\documentclass[preprint,12pt]{elsarticle}

\usepackage{amssymb}
\usepackage{nomencl}
\usepackage{url}
\usepackage{hyperref}
\usepackage{xcolor}
\usepackage{array}
\usepackage{xcolor}
\usepackage{multirow}
\usepackage{multicol}
\RequirePackage{booktabs}

\journal{Acta Astronautica}

\begin{document}

\begin{frontmatter}

\title{Orbital configurations of spaceborne interferometers for studying photon rings of supermassive black holes}

\author[a]{Ben~Hudson\corref{cor1}}
\ead{bhudson@kispe.co.uk}
\author[b,c]{Leonid~I.~Gurvits}
\author[d]{Maciek~Wielgus}
\author[b]{Zsolt~Paragi}
\author[e]{Lei~Liu}
\author[e]{Weimin~Zheng}

\cortext[cor1]{Corresponding author}

\affiliation[a]{organization={KISPE Space Systems Ltd},
            city={Farnborough},
            country={United Kingdom}}
\affiliation[b]{organization={Joint Institute for VLBI ERIC},
            city={Dwingeloo},
            country={The Netherlands}}
\affiliation[c]{organization={Faculty of Aerospace Engineering, Delft University of Technology},
            city={Delft},
            country={The Netherlands}}
\affiliation[d]{organization={Max-Planck-Institut fuer Radioastronomie},
            city={Bonn},
            country={Germany}}
\affiliation[e]{organization={Shanghai Astronomical Observatory, Chinese Academy of Sciences},
            city={Shanghai},
            country={China}}

\begin{abstract}
Recent advances in technology coupled with the progress of observational radio astronomy methods resulted in achieving a major milestone of astrophysics - a direct image of the shadow of a supermassive black hole, taken by the Earth-based Event Horizon Telescope (EHT). The EHT was able to achieve a resolution of \(\sim\)20 \(\mu\)as, enabling it to resolve the shadows of the black holes in the centres of two celestial objects: the supergiant elliptical galaxy M87 and the Milky Way Galaxy. The EHT results mark the start of a new round of development of next generation Very Long Baseline Interferometers (VLBI) which will be able to operate at millimetre and sub-millimetre wavelengths. The inclusion of baselines exceeding the diameter of the Earth and observation at as short a wavelength as possible is imperative for further development of high resolution astronomical observations. This can be achieved by a spaceborne VLBI system. We consider the preliminary mission design of such a system, specifically focused on the detection and analysis of photon rings, an intrinsic feature of supermassive black holes. Optimised Earth, Sun-Earth L2 and Earth-Moon L2 orbit configurations for the space interferometer system are presented, all of which provide an order of magnitude improvement in resolution compared to the EHT. Such a space-borne interferometer would be able to conduct a comprehensive survey of supermassive black holes in active galactic nuclei and enable uniquely robust and accurate tests of strong gravity, through detection of the photon ring features.
\end{abstract}


\begin{keyword}
VLBI \sep Radio Interferometry \sep Spaceborne Astrophysics \sep Super Massive Black Holes \sep Photon Rings
\end{keyword}

\end{frontmatter}

\makenomenclature
\nomenclature{$ALMA$}{Atacama Large millimetre Array}
\nomenclature{$AGN$}{Active Galactic Nuclei}
\nomenclature{$VLBI$}{Very Long Baseline Interferometry}
\nomenclature{$EHT$}{Event Horizon Telescope}
\nomenclature{$CR3BP$}{Circular Restricted Three Body Problem}
\nomenclature{$THEZA$}{TeraHertz Exploration and Zooming-in for Astrophysics}
\nomenclature{$SMBH$}{Super Massive Black Hole}
\nomenclature{$EHI$}{Event Horizon Imager}
\nomenclature{$LEO$}{Low Earth Orbit}
\nomenclature{$GEO$}{Geostationary Earth Orbit}
\nomenclature{$HEO$}{Highly Elliptical Orbit}
\nomenclature{$ISL$}{Inter Satellite Link}
\nomenclature{$MEO$}{Medium Earth Orbit}
\nomenclature{$ngEHT$}{Next Generation EHT}
\nomenclature{$GRMHD$}{General Relativistic Magnetohydrodynamics}
\nomenclature{$SNR$}{Signal-to-Noise Ratio}
\nomenclature{$VSOP$}{VLBI Space Observatory Programme}
\nomenclature{$JWST$}{James Webb Space Telescope}
\nomenclature{$SIS$}{Superconductor-Insulator-Superconductor}
\nomenclature{$HEB$}{Hot Electron Bolometer}
\nomenclature{$NRHO$}{Near Rectilinear Halo Orbit}
\nomenclature{$LLCD$}{Lunar Laser Communications Demonstration}
\nomenclature{$GR$}{General Relativity}
\nomenclature{$Sgr\,A^*$}{Sagittarius A*}
\nomenclature{$EHE$}{Event Horizon Explorer}
\nomenclature{$SALTUS$}{Single Aperture Large Telescope for Universe Studies}
\nomenclature{$SEFD$}{System Equivalent Flux Density}
\printnomenclature

\section{Introduction}
\label{s:intro}

\noindent Very Long Baseline Interferometry (VLBI) achieves the sharpest angular resolution of all astronomical observational techniques by combining the signals from a single source received by multiple antennas to generate a virtual telescope with a very large diameter. The science of black holes and more specifically, supermassive black holes (SMBH) in active galactic nuclei (AGN) has witnessed tremendous progress in recent years. Owing to ALMA, a connected radio interferometer operating at millimetre and submillimetre wavelengths and consisting of 66 telescopes, and subsequently the Event Horizon Telescope (EHT), sensitivity and angular resolution of studies of the innermost areas of AGN have been greatly improved \cite{combes_alma_2014, combes_alma_2019}. In 2019 and 2022 respectively, the EHT presented the first images of shadows of the SMBH M87* and the black hole at the Galactic Centre, Sgr\,A* with a resolution of around 20~\(\mu\)as at a wavelength of 1.3~mm \cite{collaboration_first_2019-2,collaboration_first_2022}.

The EHT images of M87* and Sgr\,A* show photons with bent paths near the SMBH. Besides gravity, the variable accretion environment has a significant impact on the final appearance of the images at any given time. On the other hand, images of photons that experience a complete turn, or multiple turns around a black hole appear as a series of very narrow, nearly circular structures we call \emph{photon rings}, the fine structure of which is purely determined by black hole properties and its impact on spacetime curvature. These photon ring substructures are completely smeared out in EHT images because of insufficient resolution.

While VLBI is well established at centimetre and millimetre wavelengths on the Earth, expanding VLBI arrays with space-based antennas can provide significant advantages by drastically improving the angular resolution and, for specially designed orbital configurations, \emph{(u,v)} coverage.

TeraHertz Exploration and Zooming-in for Astrophysics (THEZA) is a concept of such a system, prepared in response to ESA’s call for its next science program Voyage~2050 \cite{gurvits_theza_2021,gurvits_science_2022}. THEZA’s goal is to achieve at least an order of magnitude improvement on the angular resolution of the Earth-based EHT by observing at sub-mm wavelengths and longer baselines. In this paper the THEZA concept is explored in more depth, focusing on the practical implementation of a space-based interferometer with the specific aim of detecting and analysing the pattern of photon rings, forming in an image of a black hole. Whilst \cite{gurvits_theza_2021} presented the initial THEZA concept proposal and \cite{gurvits_science_2022} discussed the specific science case and challenges of such a mission, here we analyse a range of suitable orbital configurations for THEZA. This analysis focuses on the design of the orbits for the task of photon ring detection and includes discussion of the mission and system trade-offs that will have to be resolved as part of this process.

Although image reconstruction from VLBI data is discussed briefly in section \ref{ss:discussImaging} in the context of THEZA, the study of photon rings does not necessarily require this. Sampling of the black hole's interferometric signature with sufficient sensitivity and resolution can provide the data required to characterise the photon ring structures through focused modelling, without having to perform image reconstruction. Johnson et al. \cite{johnson_universal_2020}, Gralla, Lupsasca and Marrone \cite{gralla_shape_2020} and Broderick et al. \cite{broderick_measuring_2022} predict that the interferometric signatures of a SMBH's photon rings can provide an approach for constraining the mass and spin of black holes and for conducting consistency tests of general relatively (GR), using only a sparse interferometer array. Wielgus et al. \cite{wielgus_photon_2021} propose general tests of the Kerr metric, beyond GR, with photon ring measurements. These methods would enable a space-based VLBI array of only two elements to conduct pioneering black hole and strong gravity research.

Section \ref{ss:VLBI} provides a brief overview of the VLBI technique. Section \ref{ss:science} describes the photon ring structures and how they can be detected on very long baselines. Section \ref{s:system} presents the architecture of this specific example implementation of THEZA and provides the assumptions that have been made with respect to the interferometer's design and operation. Section \ref{s:photonRingDetection} expands on the requirement of a detection of the photon rings with the THEZA system. Three potential orbit configurations of THEZA are presented in section \ref{s:orbitConfigs}, each of which has been evaluated in terms of its feasibility and suitability for performing photon ring detection. Section \ref{ss:discussPhoton} summarises the outcomes from the orbit selection and the preliminary mission analysis performed. Section \ref{ss:discussImaging} includes a brief discussion of how THEZA could also operate in concordance with ground-based arrays for general image reconstruction of target sources. Finally, in section \ref{ss:designVariations} a number of promising future technologies are presented along with a discussion on how these could impact observations.

\subsection{VLBI on the Ground and in Space}
\label{ss:VLBI}

\noindent VLBI uses the principles of interferometry to achieve angular resolutions higher than alternative observational techniques in astronomy. A simple two element interferometer has a diffraction-limited angular resolution $\theta$ given by $\theta \approx \lambda/D$, where $\lambda$ is the wavelength and  $D$ is the distance between the two antennas, projected on the plane perpendicular to the direction to the source. This makes it insensitive to structures in the source significantly larger than \(\lambda/D\) and only provides instantaneous spatial information along one dimension of the target. Multi-telescope VLBI further improves upon these results by combining the measurements from multiple antennas in a process known as correlation. This allows the telescope array to sample different spatial scales in the source and many orientations.

The output of the VLBI correlator provides measurements of the visibility function which represents the Fourier transform of the brightness distribution in the source (see \cite{TMS-2017} for a detailed description). These measurements are obtained at spatial frequencies $u$ and $v$, defining the Fourier domain of the image, which is itself given as a brightness distribution on sky. A set of \emph{(u,v)} points at which measurements of the visibility function are obtained is called a \emph{(u,v)} coverage. It depends on the geometry of the interferometer and time as the elements of the interferometer move with respect to the target source due to the Earth rotation, for a ground element. Optimising \emph{(u,v)} coverage by increasing the density of the sampling of the plane and doing so across a wide range of baselines is a key objective for VLBI observations, allowing the highest possible fidelity of the reconstructed source model to be achieved.

Ground-based VLBI arrays have the maximum baseline length limited by the Earth diameter. They are also effectively limited to observations at frequencies below $\sim350$~GHz due to atmospheric absorption above this frequency. Both of these effects mean that the angular resolution achievable by Earth-based systems is inherently limited, and its sharpest values have already been de facto demonstrated by the EHT. The benefits of extending VLBI arrays to space have long been known and past missions such as VSOP and RadioAstron have demonstrated this on Earth-space baselines, at frequencies below 30~GHz \cite{Hirax+1998Sci, kardashev_radioastron-telescope_2013, gurvits_space_2020}. 

Extending VLBI arrays into space can drastically increase the length of the baselines and provide more diversified (in some cases -- rapid) \emph{(u,v)} coverage as the motion of the baseline is no longer limited by the solid body rotation rate of the Earth. The atmospheric limit on observational frequency does however still apply to the Earth-space baselines. Conducting VLBI on purely space-space baselines removes this limitation and the potential of such systems has been discussed in a number of publications (see \cite{palumbo_metrics_2019, roelofs_simulations_2019, An+2020, Fish+2020, Linz+2020}). 

Conducting space-based VLBI is however an inherently challenging mission concept. Requirements such as: high accuracy determination of the baseline vectors, data handling, thermal control of the payload electronics near absolute zero and antenna manufacture and deployment are just some of the difficulties which must be overcome. As a conceptual study focusing on potential orbit configurations of a THEZA system, these technical challenges are briefly summarised here, and the reader is referred to Gurvits et al. \cite{gurvits_science_2022} for a more detailed discussion.

\subsection{Science Case}
\label{ss:science}

\noindent Photons can experience extreme deflection due to the immense gravitational force in the vicinity of a black hole. Depending on the impact parameter (angular momentum) of the photon with respect to the black hole centre, photons can complete different numbers of half-orbits, \(n\), around the black hole on their way from their source to an observer. Therefore, the photons completing \(n\) number of half-orbits produce an image that is a delayed, scaled and rotated snapshot of the universe, as observed from the perspective of the black hole, forming a feature called a photon ring. Defined by their half-orbit number \(n\), each sub-ring after the \(n=0\) direct emission appears as a sharper and brighter, but less luminous in terms of the total flux density, feature as the photons have completed more orbits about the black hole before escaping \cite{johnson_universal_2020}.

The impact of the spacetime curvature on the exact shape and diameter of consecutive subrings becomes dominant as they approach the critical curve (itself only depending on the spacetime geometry \cite{gralla_black_2019}), while the astrophysical factors, e.g., related to geometry and dynamics of the emission region, become increasingly less pronounced \cite{vincent_images_2022, paugnat_photon_2022, vincent_geometric_2021}. Hence, while gravity tests with the direct image require a marginalisation over a wide range of possible astrophysical scenarios \cite{collaboration_first_2019-3,the_event_horizon_telescope_collaboration_first_2022-1}, far more robust tests can be constructed with photon rings already for \(n=1\) \cite{broderick_measuring_2022,wielgus_photon_2021}, and probing \(n=2\) could provide the most accurate tests of strong gravity to date \cite{gralla_shape_2020,paugnat_photon_2022}.

\begin{figure*}[t]
  \centering  \includegraphics[width=\columnwidth]{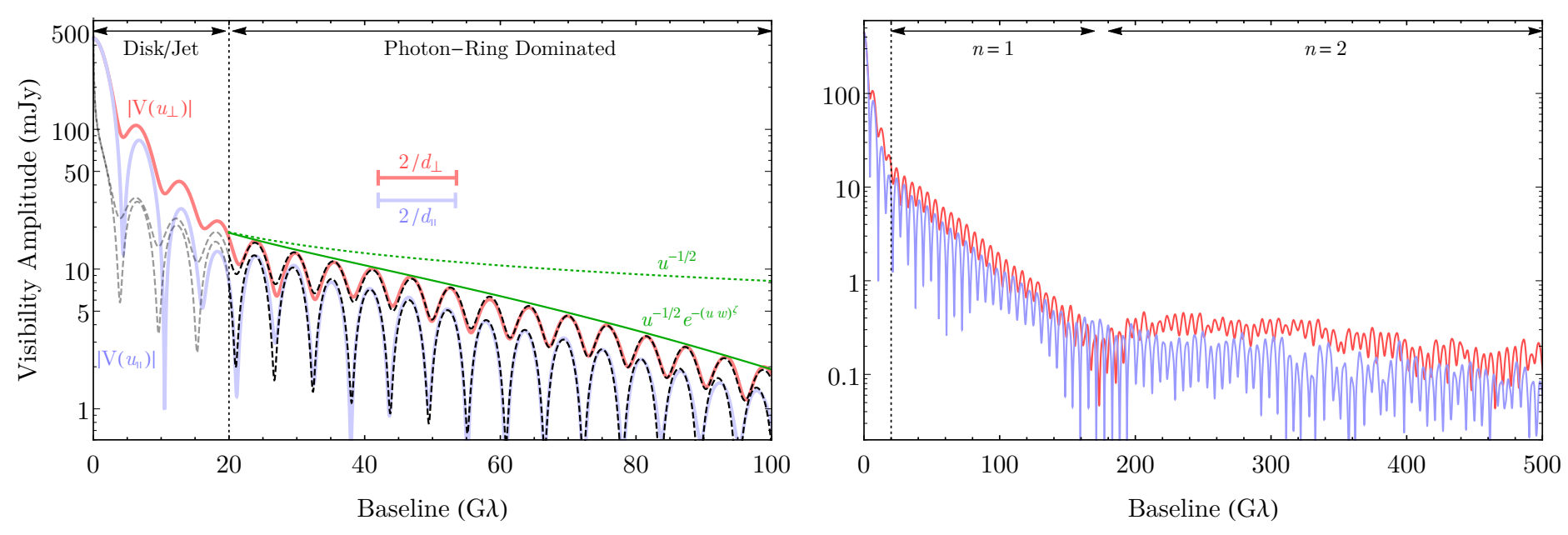}
  \caption{Figure adopted from Johnson et al. \cite{johnson_universal_2020}. \newline Left: Time-averaged interferometric signature of a GRMHD black hole model using parameters consistent with the 2017 EHT observations of M87*. Visibility amplitude shown for baselines perpendicular (red) and parallel (blue) to the black hole spin axis. The period of the amplitude variation can be used to calculate the photon ring diameters. The solid green curve envelopes the damped cascade of oscillations and carries information about the ring thickness. The grey-dashed ine depicts a simple model of the ring signature, defined in Fig. 4 of Johnson et al. \cite{johnson_universal_2020}. \newline Right: Interferometric signature on longer baselines illustrating the regions in which the first \(n=1\) and second \(n=2\) order rings dominate the total signal.}
  \label{f:THEZAphoton}
\end{figure*}

GRMHD simulations and analytic estimates suggest that the photon rings provide only \(\sim\)10\% of the total image flux density of a black hole's immediate vicinity. However, Johnson et al. \cite{johnson_universal_2020} demonstrate that the interferometric signature of a black hole on very long baselines (i.e., the most compact structural component of the image) is dominated by the photon ring contribution. Fig.~\ref{f:THEZAphoton} illustrates the visibility amplitude principally obtainable in VLBI measurements of M87*, modelled with GRMHD simulations in this synthetic data test. The signal contribution is broken down into the direct emission and the individual photon subrings. Johnson et al. \cite{johnson_universal_2020} in particular demonstrate how measurements of the photon ring angular diameters can be conducted using the upper envelope of the visibility amplitude and related methods of fitting to very high spatial frequency visibility amplitudes were demonstrated by \cite{gralla_shape_2020} and \cite{paugnat_photon_2022}. Photon rings offer new and very powerful means of studying fundamental gravitational physics, astrophysics and cosmology (see \cite{johnson_universal_2020, gralla_shape_2020, broderick_measuring_2022, wielgus_photon_2021}).

A prospective sub-millimetre spaceborne interferometer will have to address multiple scientific rationales. It is reasonable to expect, and it is confirmed by the experience of the two dedicated space VLBI missions VSOP and RadioAstron completed to date, that the final set of mission specifications will be a result of trade-offs between various and often conflicting requirements (see \cite{gurvits_space_2020} and references therein). Measurements of the parameters of photon rings of high orders as discussed above can be conducted only by means of space VLBI. Thus we adopt photon ring investigation as a defining science case in the analysis of a prospective space VLBI mission. We note that as shown in THEZA studies \cite{gurvits_science_2022}, mission specifications aimed at addressing the photon ring science case are consistent with many other tasks of the next generation space VLBI missions.

\section{System Architecture and Design}
\label{s:system}

\noindent For this example implementation of the THEZA concept, we consider the following architecture. The interferometer consists of two space-based elements, each with an antenna of diameter 15~m. This choice of antenna geometry is simply postulated at this stage and a discussion into the effects of varying antenna sizes is provided in section \ref{ss:designVariations}. A primary observational frequency of 690~GHz is chosen due to its attractiveness for a variety of radio astronomy studies of cosmic molecular spectral lines and therefore advanced developments of the receivers. We note that this frequency band is considered in similar studies for space-based VLBI systems like the Event Horizon Imager (EHI) \cite{roelofs_simulations_2019} and CAPELLA \cite{CAPELLA-2023arXiv}. A range of receiver and timing characteristics are considered in section \ref{s:photonRingDetection} whilst evaluating the feasibility of photon ring detection.

We note that the simplifications made in these assumptions have significant implications. In practice, the deployment of a 15~m antenna for observation at 690~GHz would require a surface accuracy on the order of 10~\(\mu\)m which is far beyond the current capabilities of antenna manufacture. The potential use of a phased array configuration is discussed in Gurvits et al. as an alternative to this and the reader is referred to that text for more information \cite{gurvits_science_2022}. In addition, performance as a function of the observational frequency is itself a trade-off as the Signal-Noise Ratio (SNR) is degraded at higher frequencies due to antenna surface deficiencies, receiver noise and the limits of timing accuracy. These effects must be weighed against the improvements in resolution and scattering cross-sections that higher frequencies provide.

Other potential features of the spacecraft design are discussed briefly in the following analysis. Thermal control techniques are considered including the impact on operations if a Sun shield is implemented and the utilisation of cryogenic cooling of the receiver electronics. Possible solutions to the communications challenges faced by a space-based VLBI mission (e.g. optical communications systems and Inter-Satellite Links (ISLs)) are also discussed. However, detailed design of the system beyond high-level consideration of these features is outside of the scope of this paper and should be tackled in future work.

\section{Detecting the Photon Rings}
\label{s:photonRingDetection}

\begin{table}[t]
    \centering
      \caption{Physical baseline length associated with 20-200 G\(\lambda\) range for different frequencies.}
      \begin{tabular}{m{2cm} m{2cm} m{4cm}}
        \toprule
        Frequency \newline [GHz] & Wavelength \newline [mm] & Baseline Length \newline 20 G$\lambda$ [x$10^4$ km] \newline 200 G$\lambda$ [x$10^5$ km]\\
        \midrule
        230	& 1.30  &   2.61\\
        345	& 0.87  &   1.74\\
        590	& 0.51  &   1.02\\
        690	& 0.43  &   0.87\\
        1200 & 0.25 &   0.50\\
        \bottomrule
    \end{tabular}
    \label{tab:THEZA_Photon}
\end{table}

\noindent For the current study, M87* and Sgr\,A* are chosen as the main sources when demonstrating \emph{(u,v)} coverage and sampling of the visibility amplitude achieved by different THEZA configurations. Each interferometer configuration's ability to observe a variety of sources across the celestial sphere is also evaluated, ensuring that the system is highly versatile.

As can be seen in Fig.~\ref{f:THEZAphoton}, the time-averaged interferometric signature is dominated by the first photon ring at baselines longer than 20~G\(\lambda\). The second order photon ring \(n=2\) starts to be the dominant contributor to the total signal beyond 160~G\(\lambda\). Because of this, a THEZA mission achieving a baseline variation of between 20 and 200~G\(\lambda\) (\(\sim\)10 and 1~\(\mu\)as) would be theoretically capable of measuring the interferometric signature of the \(n=1\) and \(n=2\) photon rings. This would enable the most robust to date, mass/distance prior independent tests based on the rings' diameter ratio \cite{broderick_measuring_2022,wielgus_photon_2021}.

The physical length of the baseline between 20 and 200~G\(\lambda\) depends on the observational frequency of the system and thus, so does the specific orbital geometry required, as presented in Table~\ref{tab:THEZA_Photon}. The variation in the proposed orbit configurations' geometries is therefore shown for a range of frequencies up to 1200~GHz as the upper bound on THEZA's capabilities \cite{gurvits_science_2022}. Frequencies of 230 and 345~GHz are used by the ground-based EHT \cite{collaboration_first_2019} and the proposed ngEHT \cite{blackburn_studying_2019,johnson_key_2023}, and are considered as the frequency bands for analysing the potential of the THEZA system to be used in combination with ground-based arrays for imaging VLBI experiments. The additional benefit of increasing the observing frequency above 230~GHz is the related decrease of the optical depth, and hence larger fractional contribution of the photon ring to the total source flux density \cite{vincent_images_2022}. Increased observational frequency also has the advantage of reducing the impact of interstellar scattering which is a concern for Sgr\,A* \cite{johnson_universal_2020}.

\begin{table}
    \centering
     \caption{Variation in \(\sigma_{thermal}\) [mJy] for observing with a range of integration time and bandwidth combinations for a baseline between two  15~m antennas.}
    \begin{tabular}{m{2.0cm} m{1.5cm}  m{1.5cm}  m{1.5cm}  m{1.5cm}}
        \toprule
        & \multicolumn{4}{c}{Bandwidth [GHz]} \\
        \midrule
        $t_{int}$ [s] & 4 & 16 & 32 & 64 \\
        \midrule
        1 & 60.8 & 30.4 & 21.5 & 15.2 \\
        10 & 19.2 & 9.62 & 6.80 & 4.81 \\
        100 & 6.08 & 3.04 & 2.15 & 1.52 \\
        1000 & 1.92 & 0.96 & 0.68 & 0.48 \\
        \bottomrule
    \end{tabular}
    \label{tab:Thermal}
\end{table}

Minimising the thermal noise of the interferometer system will be essential for detecting the weak signals of the photon rings on longer baselines. The effect of varying integration times and bandwidths on the thermal noise is shown in Table \ref{tab:Thermal} for an example 15~m antenna. The EHT currently observes with a total bandwidth of 8~GHz \cite{collaboration_first_2019}, with future upgrade considerations of up to 4x8~GHz \cite{blackburn_studying_2019}. Here bandwidths of 4-64~GHz are considered for the sensitivity calculations. The higher end of the integration time range of 1-1000~s analysed here would likely require developments in frequency standards to increase clock stability as this will limit the useful integration time. Gurvits et al. provide a summary of potential frequency standard developments for spaceborne VLBI systems of the future \cite{gurvits_science_2022}. Parameters from the EHI study observing at the same 690 GHz are used in these calculations, including a system noise temperature, \(T_{sys}\) of 150 K \cite{roelofs_simulations_2019}. In order to be consistent with the previous study of the topic by Roelofs et al. \cite{roelofs_simulations_2019}, we adopt their value $\eta = 0.48$ to account for aperture, clock and correlator efficiencies. However, we note that this value will have to be re-estimated for a specific instrumental configuration of THEZA in the next steps of project development. With these values, each THEZA spacecraft has a system equivalent flux density (SEFD) of 4883~Jy, for the proposed 15~m antennas.

Considering Fig.~\ref{f:THEZAphoton}, the ring diameter is encoded in the period of the oscillations of the interferometric signature. In order to measure the diameter of the photon ring dominating the signal at a particular baseline, we need to determine the period of the visibility domain signature by confidently detecting locations of its maxima and minima. Considering this, a sensitivity of 5~mJy would provide a conservative requirement to characterise minima and maxima at 20-30~G\(\lambda\). A baseline sensitivity of \(\sim\)1~mJy would enable characterisation at \(\sim\)100~G\(\lambda\) and for baselines beyond \(\sim\)200~G\(\lambda\), a sensitivity of less than 0.2~mJy would be required to measure the angular diameter of the second order ring.

For an interferometric response detection at each \emph{(u,v)} point, we adopt an SNR of at least 5. Escalating this requirement to higher values of SNR will be addressed in separate follow-up studies. As we are primarily concerned with the visibility amplitudes of the signal here, incoherent stacking of multiple measurements at the same \emph{(u,v)} point could provide a means of reducing the required SNR to claim a detection. Stacking of \(N\) samples incoherently reduces error in the signal detection by \(\sqrt{N}\) which would allow measurements at the same \emph{(u,v)} locations to be repeated in order to reach an unambiguous detection \cite{rogers_fringe_1995}.

Based on this analysis, accurate measurement of the Fourier visibility at the spatial scales required for first order photon ring detection would be achievable when operating with an integration time and bandwidth combination of 1000~s and 32~GHz, up to a baseline length of 100~G\(\lambda\). Characterisation of the second order ring would likely require a sensitivity of \(<\) 0.2 mJy, a factor of 2 improvement over the sensitivity provided by a combination of the integration time of 1000~s and bandwidth of 64~GHz in Table \ref{tab:Thermal}. However, incoherent stacking of multiple measurements across time could provide a promising method for enabling detection of the \(n=2\) ring.

For the orbit configurations presented in section \ref{s:orbitConfigs}, to maximise the efficiency of the mission the aim is to enable a photon ring detection as quickly as possible. However, we note that the geometry of the \(n=1\) and higher order photon ring structures is stable on human timescales and therefore observation could take place over a far longer period of time than is presented here in order to enable a first detection. Considering this, where possible the configurations have been designed to achieve the required baseline variation in 7~days. This is also beneficial for imaging of the overall source as the EHT did, including \(n=0\), for observations that may take place with ground-based arrays at a lower frequency (see discussion in section \ref{ss:discussImaging} on a multiple receiver configuration of the interferometer elements). Observation over 7~days is sufficient for accurate image reconstruction of M87*. However, although the photon rings (\(n=1\) and higher) of Sgr\,A* are stable over time, the dynamic timescale of the overall source is on the order of minutes and imaging across 7~days would result in significant blurring of the image (see discussion in \cite{roelofs_simulations_2019}). Achieving the baseline variation in the dynamic timescale of Sgr\,A* is unrealistic for THEZA. The data collected from Sgr\,A* will still however be useful for other studies such as inferring mean source morphology, thus determining key parameters like the brightness temperature.

To summarise, the primary objective of the THEZA configurations presented below, and the one which takes priority in the evaluation criteria, is the ability to detect the first order photon ring (\(n=1\)). Detection of the \(n=2\) photon ring is considered as a secondary objective, as is the ability to perform observations in collaboration with ground-based arrays, operating at lower frequencies.

\section{Orbit Configurations}
\label{s:orbitConfigs}

\noindent 
The main performance requirements of THEZA for the detection and analysis of photon rings, and more general imaging in collaboration with ground-based arrays are as follows:

\begin{enumerate}
    \item Achieve a 1-dimensional, projected baseline variation between 20~G\(\lambda\) and 200~G\(\lambda\), as described in section \ref{s:photonRingDetection}
    \item Sample the \emph{(u,v)} plane at a minimum of the Nyquist frequency requirement to accurately produce the damped cascade of oscillations illustrated in Fig. \ref{f:THEZAphoton}
    \item Be capable of achieving a 1~\(\mu\)as resolution for a range of radio sources, across the celestial sphere (the primary objective of THEZA \cite{gurvits_theza_2021})
\end{enumerate}

\noindent Orbital configurations for space-based interferometers have been presented in several previous studies. Roelofs et al. considers a constellation of three spacecraft in a Medium Earth Orbit (MEO) for high-fidelity imaging of the black hole shadows in M87* and Sgr\,A*, with an order of magnitude higher resolution than achieved by the EHT \cite{roelofs_ngeht_2023}. A modification of this approach with  two to four spaceborne radio telescopes on counter-rotating orbits is considered by Rudnitskyi et al. \cite{rudnitskiy_optimal_2023}. Fish et al. propose an addition to the EHT in the form of two spacecraft operating in GEO and Highly Elliptical Earth Orbit (HEO) to resolve SMBH shadows as small as 3~\(\mu\)as \cite{Fish+2020}. The Event Horizon Explorer (EHE) is a mission concept to extend the ground based EHT with a space-based antenna \cite{kurczynski_event_2022}. The technological readiness and suitability for a NASA medium-class explorer mission is currently being assessed. The EHE would perform VLBI on ground-space baselines. Possibilities also exist in ad-hoc VLBI conducted as a secondary objective of a mission. The Single Aperture Large Telescope for Universe Studies (SALTUS) is a mid/far-infrared telescope concept utilising an inflatable antenna design \cite{kim_saltus_2022}. Discussions related to the inclusion of a multiband VLBI instrument onboard the SALTUS mission are currently ongoing. A constellation of two pairs of small VLBI-equipped satellites in LEO called CAPELLA for observations at 690~GHz is considered in \cite{CAPELLA-2023arXiv}.

In this study, three primary types of THEZA configurations have been examined: operation in Earth orbit, at the Earth-Moon (EML2) and Sun-Earth L2 Lagrange (SEL2) points. The orbit families presented are a subset of solutions that can provide the required baseline variation. The specific configurations proposed are those which offer more favourable characteristics in terms of the spacecraft operations and provide the greatest flexibility for VLBI observations. Each of the configurations is evaluated by considering the impacts the specific orbit selection would have on the design and operation of the mission. In particular, thermal control, data handling and communications are focused on as part of this evaluation as these are the elements of the system that have the most demanding requirements for a space-based VLBI mission. The mission analysis performed for each of the proposed configurations is not complete and is merely intended to provide a preliminary assessment of the suitability of a specific orbit design. Factors such as power generation, radiation exposure and the times at which observations can in practice be conducted when communications, sensor field-of-view and Sun and moon blinding are taken into account will also have to be included in trade-offs as part of the orbit selection process. These should be considered in more detailed mission concept proposals in the future.

Simulations of VLBI observations have been conducted using the OmniUV toolkit\footnote{OmniUV is a multi-purpose toolkit which is capable of simulating space and ground VLBI observation and is publicly available at: \url{https://github.com/liulei/omniuv}, accessed on 2023.02.14.} to calculate the \emph{(u,v)} coverage achieved by different interferometer orbital configurations \cite{liu_omniuv_2022}.

\subsection{Earth Orbit}
\label{ss:earth}

\begin{figure*}[t]
  \centering
  \includegraphics[width=0.6\columnwidth]{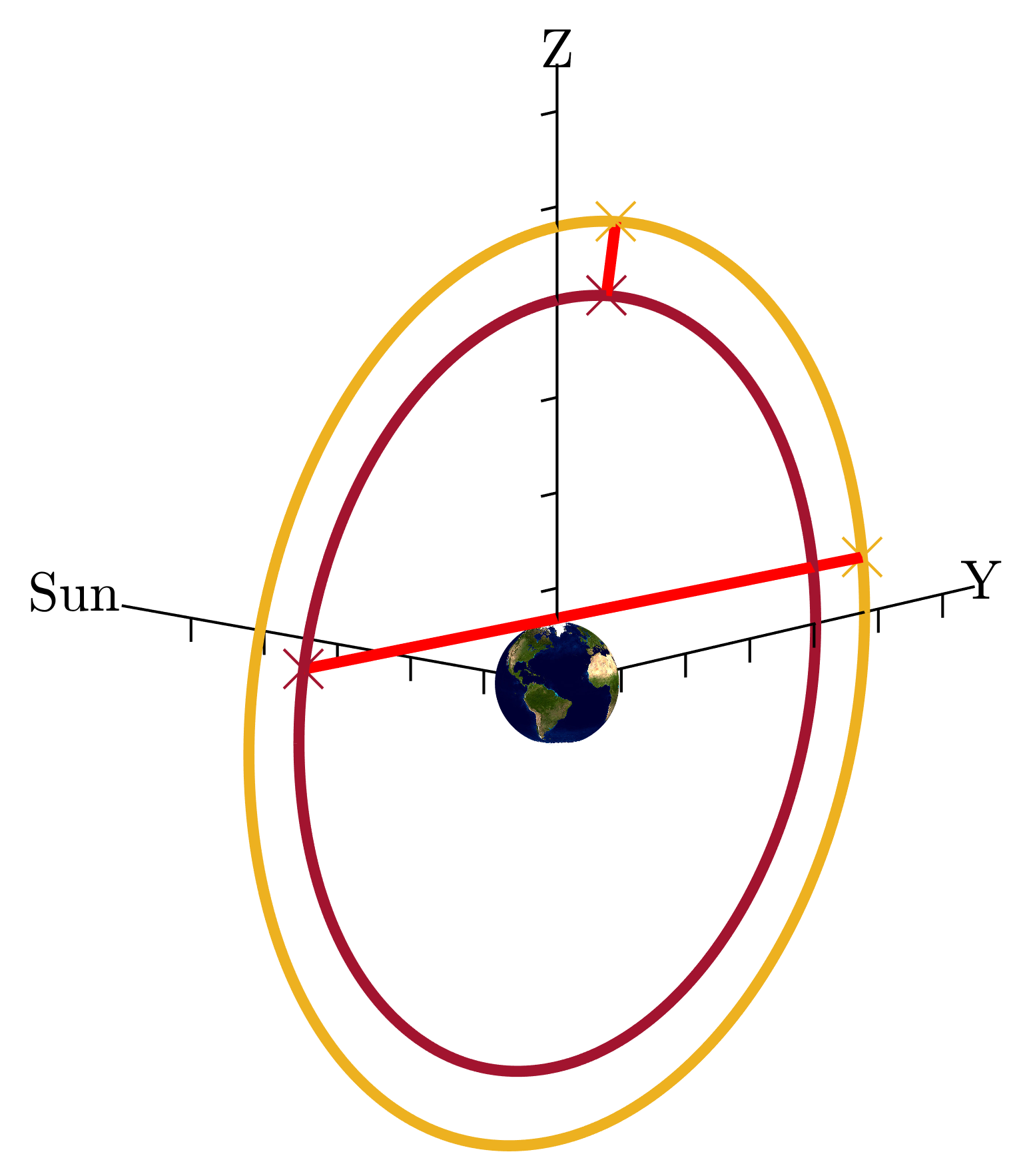}
  \caption{Two-spacecraft THEZA configuration in Earth orbit with the longest and shortest baselines illustrated in red. The configuration aims to achieve the 20--200~G\(\lambda\) baseline variation for observation at 690~GHz. Tick marks indicate 10,000~km increments.}
  \label{f:EarthOrbit}   
\end{figure*}

\noindent Depicted in Fig.~\ref{f:EarthOrbit} is a potential configuration for THEZA operating in Earth orbit, observing at 690~GHz. The orbits have been designed to achieve the baseline variation of 20--200~G\(\lambda\), required to produce the visibility amplitude plot shown in Fig. \ref{f:THEZAphoton}. The orbits are polar to maximise \emph{(u,v)} coverage of THEZA's two primary targets, M87* and Sgr\,A*, both of which are at fairly low declinations. M87* and Sgr\,A* have right ascension and declination coordinates of (187.5\(^\circ\), 12.0\(^\circ\)) and (266.4\(^\circ\), $-$29.0\(^\circ\)), respectively, to one decimal place. The two elements of the interferometer orbit in opposing directions to achieve the required baseline variation, which might require two separate launches for the interferometer elements. Variation in the right ascension of each orbit is minimised as the nodal precession effect, caused by the J2 harmonic of Earth's gravitational field, is negated by the selection of polar orbits. Although Fig. \ref{f:EarthOrbit} shows the configuration for 690~GHz, the variation of this geometry for different observational frequencies is provided in Table \ref{tab:EarthVariation}. 

\begin{table}
    \centering
    \caption{THEZA Earth orbit configuration for observations at different frequencies where THEZA 1 is the spacecraft operating in the higher altitude orbit and THEZA 2 is in the lower orbit.}
    \begin{tabular}{m{2cm} m{2cm} m{2cm} m{2cm} m{2cm}}
        \toprule
        Frequency \newline [GHz] & THEZA 1 \newline Radius [km] & THEZA 1 \newline Period [hours] & THEZA 2 \newline Radius [km] & THEZA 2 \newline Period [hours]\\
        \midrule
        230	& 144000  & 161.2 & 121000 & 125.7\\
        345	& 95800  & 90.3 & 80400 & 70.7\\
        590	& 56000  & 43.1 & 47000 & 34.1\\
        690	& 47900  & 35.0 & 40200 & 27.8\\
        1200 & 27500 & 17.2 & 23100 & 14.0\\
        \bottomrule
    \end{tabular}
    \label{tab:EarthVariation}
\end{table}

As described in Gurvits et al., the use of Superconductor - Insulator - Superconductor (SIS) mixers up to 700-800~GHz requires operation at very low temperatures (\(<\) 4~K), to maximise noise performance. Above this frequency range, Hot Electron Bolometer (HEB) mixers are required which must operate at \(\sim\) 0.3 K \cite{gurvits_science_2022}. To achieve these extremely low temperatures, cryogenic cooling of the receiver electronics would no doubt be required, very likely in combination with a Sun shield. 

To reduce the complexity of the thermal control, a strategy similar to the James Webb Space Telescope (JWST) is implemented whereby observation is always conducted away from the Sun. This keeps the Sun at a near-constant attitude with respect to the spacecraft, allowing a Sun shield to be implemented to reduce the heat absorbed from the solar flux, on average 1371 W/m\(^2\) at the Earth. Using this strategy with a minimum Sun avoidance angle of \(\sim\)85\(^\circ\) (the exact figure would of course depend on the specific Sun shield design), \(\sim\)39\% of the celestial sphere can be observed at any time with the science instruments kept in the Sun shield's shadow. This results in each source being visible for two, 60 day periods each year \cite{rigby_science_2012}. It is envisaged that a similar level of source availability would be achievable by this THEZA configuration.

\begin{figure}[t!]
  \centering
  \includegraphics[width=0.7\columnwidth]{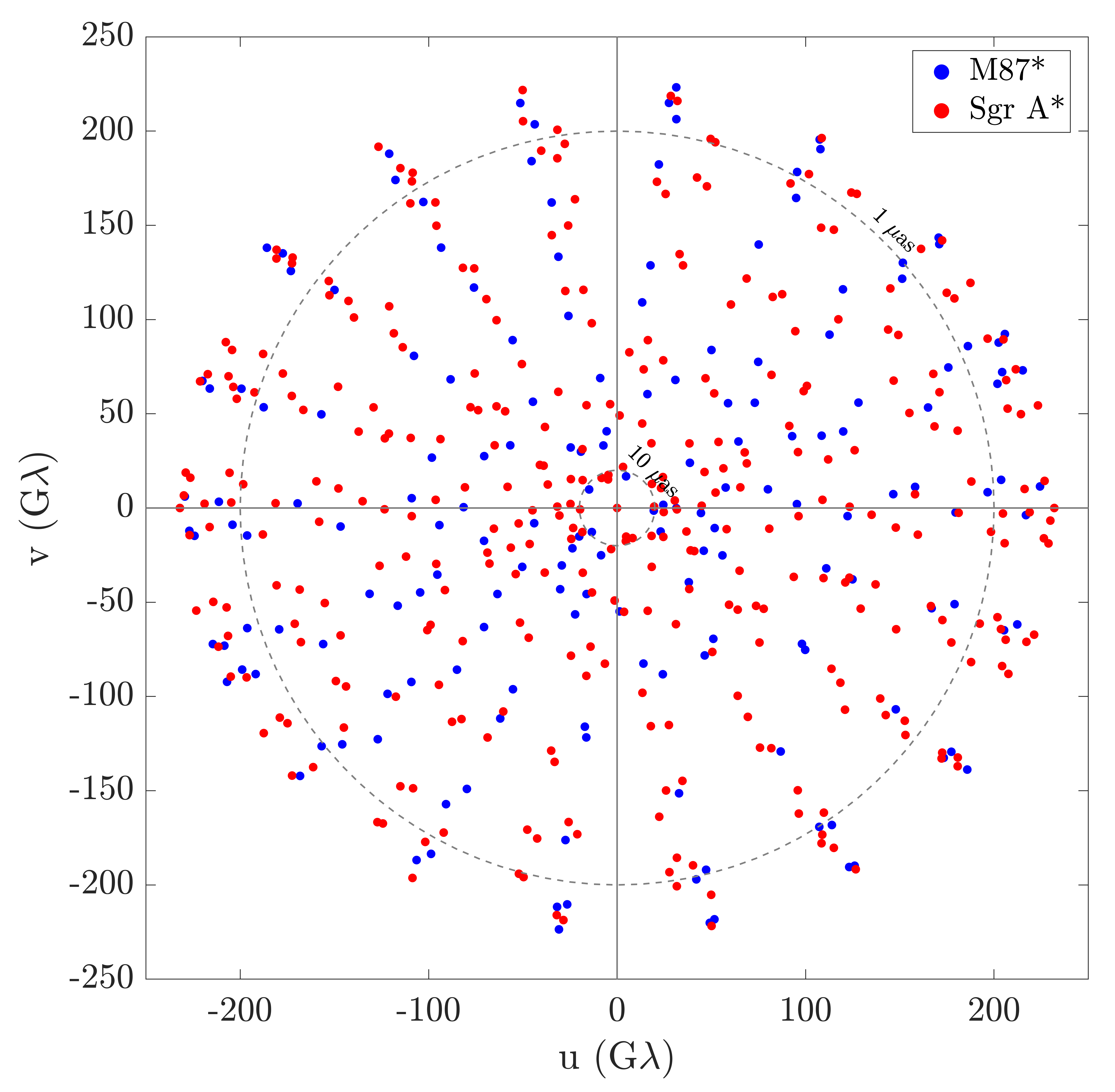}
  \caption{\emph{(u,v)} coverage of M87* and Sgr\,A* achieved by Earth orbiting THEZA interferometer across a period of 7~days. An instrument sampling cadence of 8400~s is used to achieve the required Nyquist frequency sampling of the interferometric signature.}
  \label{f:EarthUV}   
\end{figure}

Calculation of \emph{(u,v)} coverage is conducted for a maximum observation period of 7 days, as stated in section \ref{s:photonRingDetection}. Fig.~\ref{f:EarthUV} illustrates the \emph{(u,v)} coverage of THEZA for observing M87* and Sgr\,A*, when the right ascension of each source is perpendicular to the orbital plane. This offers the most extensive \emph{(u,v)} coverage that can be achieved during a year. In order to maintain a near-constant Sun attitude with respect to the spacecraft, the \emph{(u,v)} coverage becomes less extensive as the position of the source reduces the angle between its direction and the orbital plane. This is a disadvantage for image reconstruction using data from the interferometer array but the configuration still achieves the 20--200~G\(\lambda\) baseline variation required for photon ring detection, regardless of the position of the source with respect to the orbital plane. In this figure and all subsequent \emph{(u,v)} plots, each \emph{(u,v)} point represents the midpoint of an observation.

To accurately reproduce the damped cascade of visibility amplitude oscillations in Fig. \ref{f:THEZAphoton}, the interferometric signature must be sampled at a minimum of the Nyquist sampling requirement. Here we consider the \emph{cadence} of observations to mean the time between two successive sets of sampling by the interferometer. Fig. \ref{f:EarthUV} depicts the coverage from a simulation of THEZA conducting observations at this minimum rate. The \emph{(u,v)} coverage can of course be increased by observing at a higher cadence and/or in conjunction with ground-based antennas at lower frequencies. The period of the oscillations in Fig. \ref{f:THEZAphoton} is \(\sim\)5~G\(\lambda\) resulting in a Nyquist sampling requirement of 2.5~G\(\lambda\). Observing over 7 days results in a observation cadence of 8400~s. Other factors affecting the achievable cadence of the instrument will include the data storage capabilities of the mission, thermal considerations and power requirements. Although such factors cannot yet be assessed at this early stage, a cadence far longer than the integration time of the instrument (in this case 1000~s) is more likely to be achievable in reality.

\begin{figure}
  \centering
  \includegraphics[width=0.9\columnwidth]{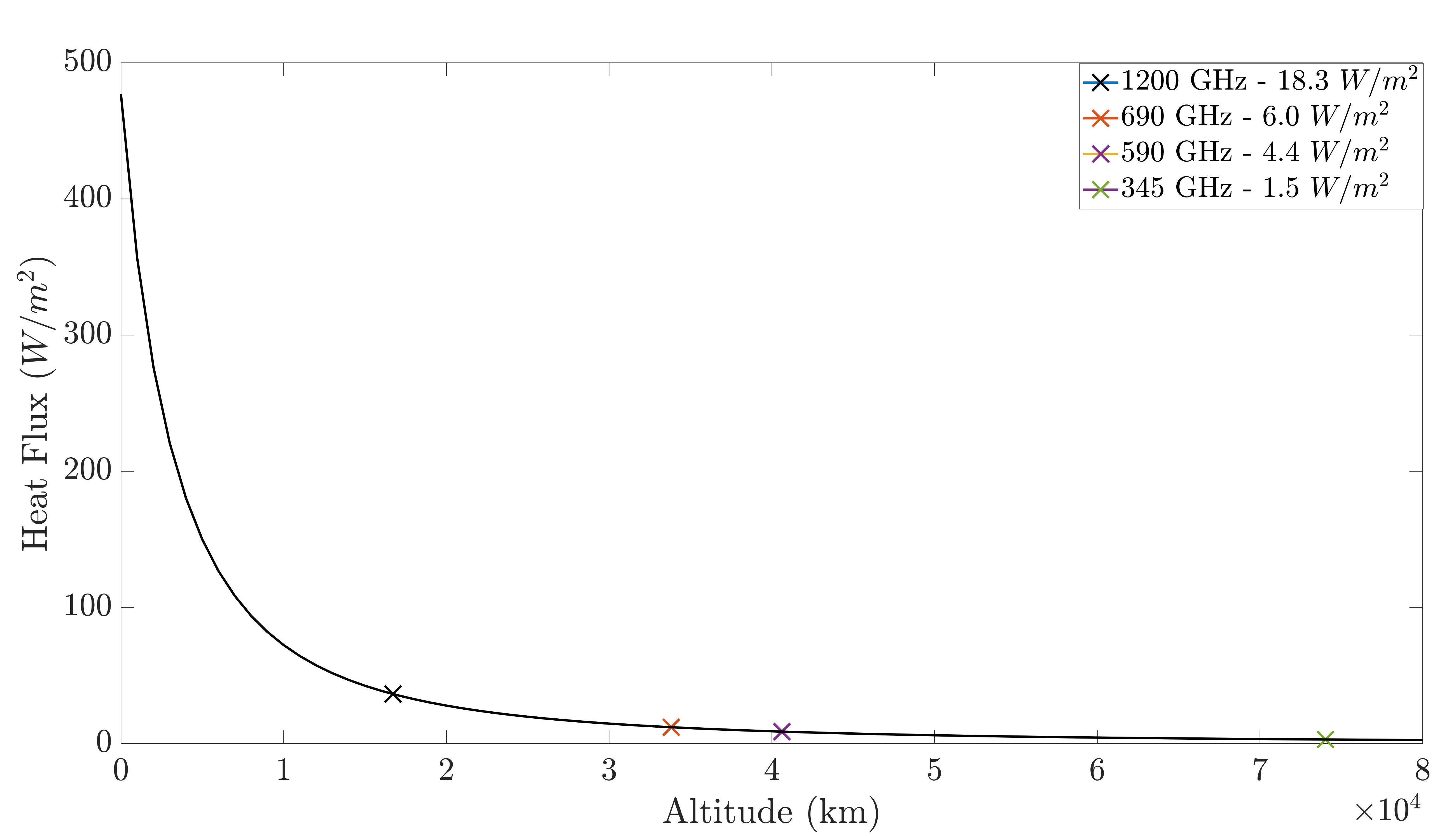}
  \caption{Heat flux experienced at different orbital altitudes including infrared thermal emission from the Earth and the albedo effect, both of which vary as inverse square relationships with altitude.}
  \label{f:EarthHeatFlux}   
\end{figure}

Although the Sun direction can be effectively shielded, the Earth itself produces a significant thermal emission, primarily in the infrared band. There is also the albedo effect, solar energy reflected from the Earth which contributes significantly to the total heat flux. Fig.~\ref{f:EarthHeatFlux} depicts the relationship between the total heat flux and altitude, assuming an albedo coefficient of 0.34 and Earth thermal emission of 240 W/m\(^2\). For observation at 690 GHz, the total Earth emission at the lowest spacecraft is only 6.0 W/m\(^2\), significantly less than the solar flux from the Sun.

VLBI requires operation at wide bandwidths to achieve sufficient sensitivity. For a 24~hour observation, the amount of VLBI data might reach terabytes per interferometer element. The EHT has been operating at 64~Gbps since 2018 \cite{collaboration_first_2019}. A real-time downlink technology could be implemented to remove the requirement for a very large data storage system onboard. This does however impact the availability of the system to conduct observations as a ground station must be in contact with each of the two interferometer elements. Development of mass data storage systems, suitable for operation in space would allow observations to take place outside of ground station contacts and downlink the data at a later time. In the context of VLBI, such a data storage system would not necessarily require radiation hardening of the memory itself as radiation-induced bit flips are tolerable in VLBI data, even up to several percent of the overall number of data points. Radiation tolerance would however be a problem for components susceptible to single-point failures such as memory management logic. 

Optical communications systems such as that used by NASA's TBIRD mission have recently demonstrated burst communications up to 200~Gbps from Low Earth Orbit (LEO) \cite{tavares_nasa_2023}. A VLBI-oriented study of a data downlink system from GEO with the data rate exceeding 256\,Gbps has been presented recently in connection with the EHE concept \cite{Wang+2023SPIE}. It should be noted that the TBIRD downlink duty cycle is limited due to the need to be within line-of-sight of an appropriate ground station as would be expected for a LEO spacecraft. Conversely, optical communications from GEO can downlink continuously at lower rates. This highlights the need for the access time between the interferometer elements and optical communication ground stations to be to be taken into account when analysing the downlink capabilities of the mission. The locations of optical communication ground stations is also limited by the need for high elevation, low humidity sites. However, considering the current rate of development, it is reasonable to envisage that future systems will be capable of achieving the rate required to downlink data from the proposed orbits for THEZA.

\subsection{Earth-Moon L2}
\label{ss:eml2}

\begin{figure*}[t!]
  \centering
  \includegraphics[width=0.6\columnwidth]{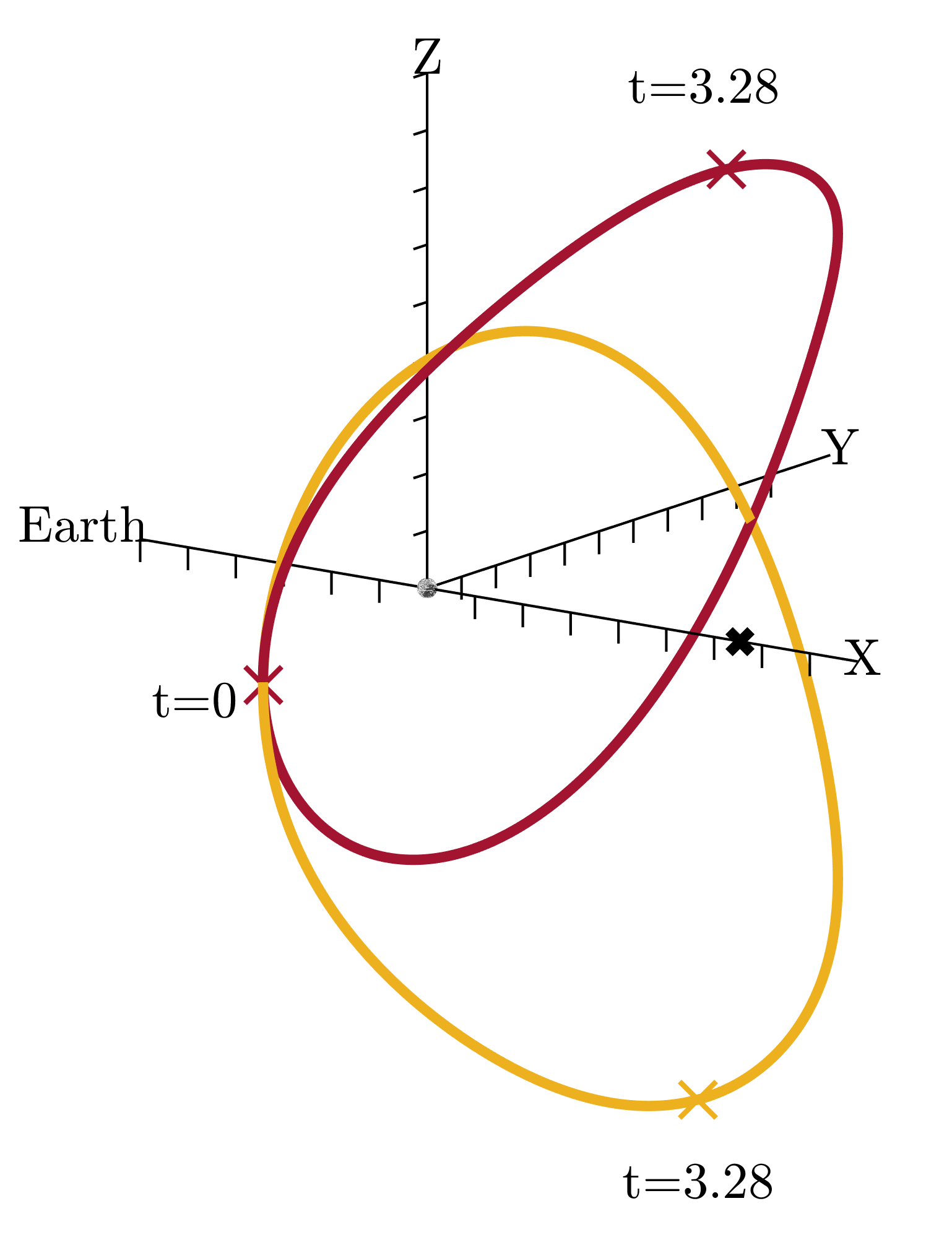}
  \caption{THEZA configuration in Halo orbits about EML2 observing at 690~GHz. Moon-centred coordinate frame with Earth direction depicted. The position of the L2 point is shown by a cross. Position of interferometer elements throughout orbit illustrated with time given in days. Tick marks indicate 10,000~km increments.}
  \label{f:EML2Orbit}   
\end{figure*}

\noindent The Lagrange (or libration) points are positions of equilibrium in the circular restricted three-body problem (CR3BP). Families of periodic orbits exist about these points which require very little station-keeping in order to maintain the spacecraft trajectory. Of the five Lagrange positions, the L2 point is considered here as it has a number of favourable characteristics including a direct line-of-sight of the Earth. Periodic orbits about the L2 point include Lissajous, Lyapunov, Vertical, Axial and Halo types. Each of these families have different orbital geometries. For example, the Vertical orbits describe a figure-of-eight trajectory about the L2 point. For a THEZA configuration in the EML2 system, the Halo orbit types were analysed as they exhibit the shortest periods of the L2 orbit families, they never enter eclipse behind the secondary body and the Near-Rectilinear Halo Orbits (NRHO) considered here are highly stable. The Halo orbits have been modelled using Richardson's third-order approximation, a technique with sufficient accuracy for this preliminary concept study \cite{richardson_analytic_1980}. The reader is referred to Vallado for a more detailed description of the libration point periodic orbit families \cite{vallado_fundamentals_2013}.

\begin{figure}[t]
  \centering
  \includegraphics[width=0.6\columnwidth]{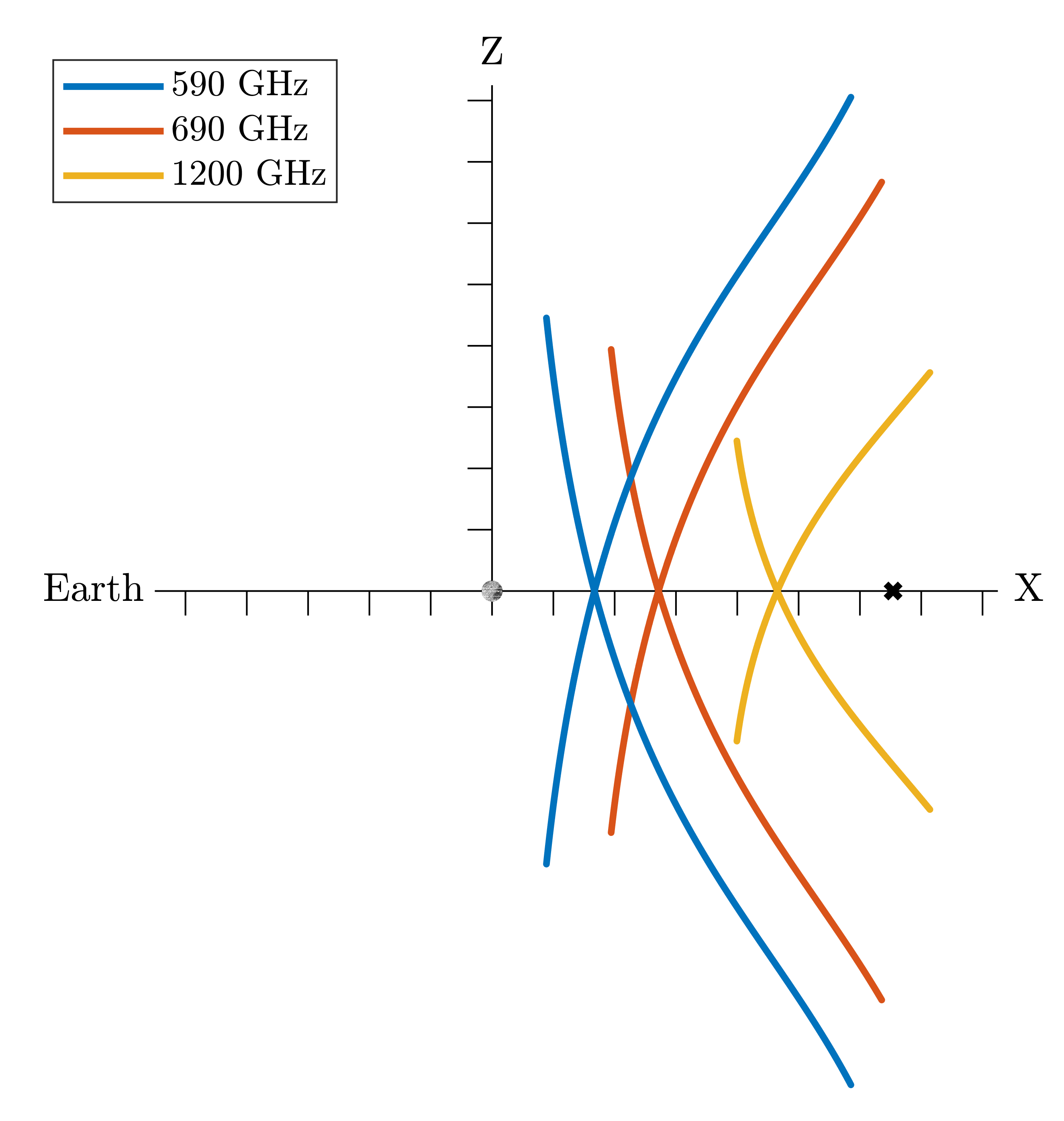}
  \caption{Variation in EML2 THEZA configuration for observing at different frequencies. The position of the L2 point is shown by a cross. Tick marks indicate 10,000~km increments.}
  \label{f:EML2Freq}   
\end{figure}

Periodic orbits about the EML2 point provide some potentially attractive alternatives to Earth-orbiting configurations of THEZA. Depicted in Fig. \ref{f:EML2Orbit} is a configuration for two THEZA spacecraft in orbit about the L2 point in the Earth-Moon system. The L2 point is located \(\sim\)61,350~km from the moon. The design consists of two identically shaped northern and southern Halo orbits. The positions of the spacecraft at different times is shown in Fig.~\ref{f:EML2Orbit} to illustrate this motion. The Halo orbits used have a $z$-excursion of 51,300~km for observing at 690~GHz. Fig.~\ref{f:EML2Freq} depicts the variation in the size of the Halo orbits for observation at different frequencies and Table \ref{tab:EML2Variation} provides the parameters for these orbits.

\begin{figure}[t!]
  \centering
  \includegraphics[width=0.9\columnwidth]{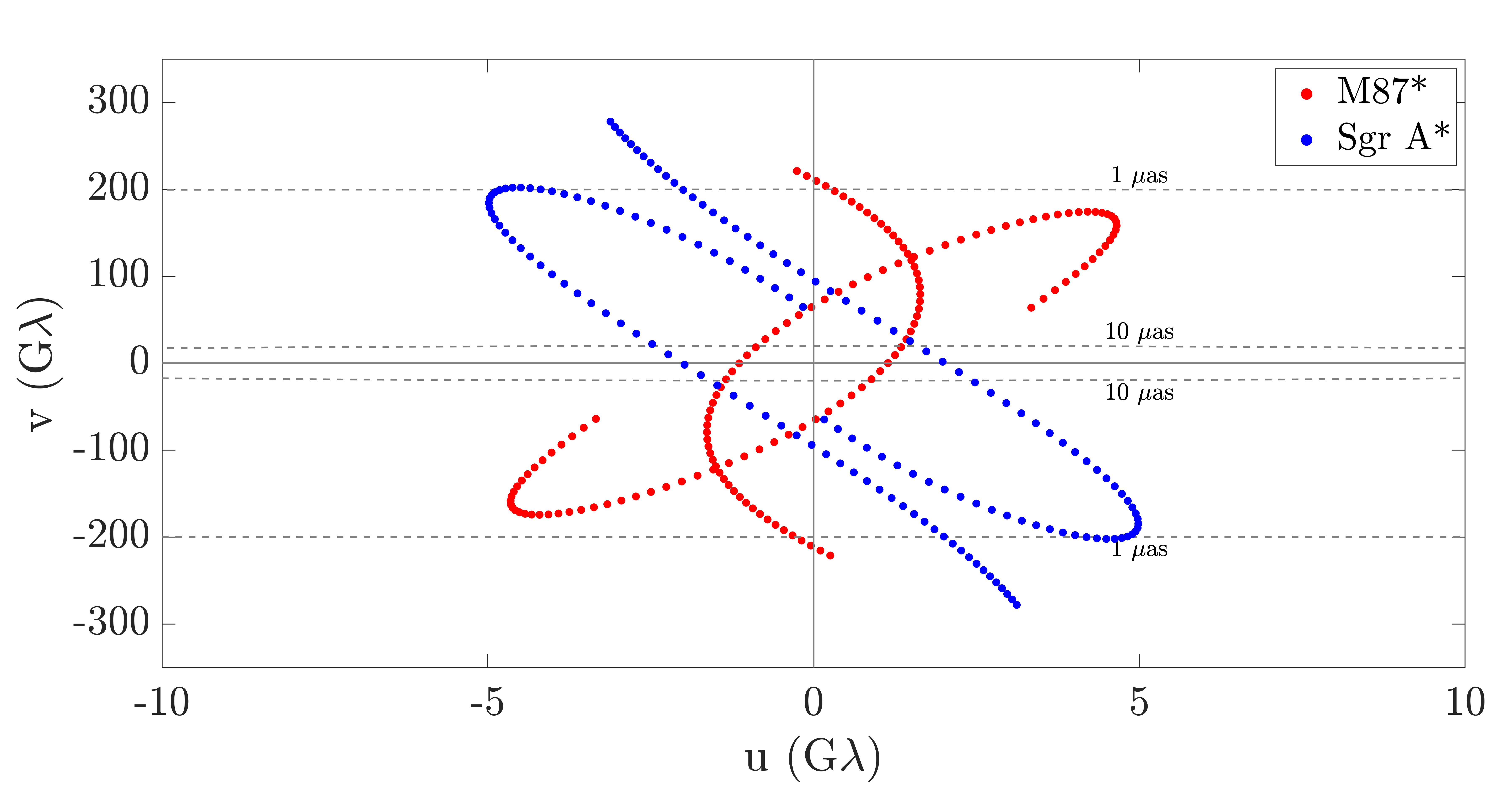}
  \caption{\emph{(u,v)} coverage of M87* and Sgr\,A* achieved by EML2 THEZA configuration across a 7~day period. An instrument sampling cadence of 7020~s is used to achieve the required Nyquist frequency sampling of the interferometric signature.}
  \label{f:EML2uv}   
\end{figure}

The \emph{(u,v)} coverage of this configuration of M87* and Sgr\,A* is depicted in Fig. \ref{f:EML2uv}. As can be seen, the required baseline variation of 20-200~G\(\lambda\) is achieved. The coverage is however extremely one-dimensional as there is no rotation of the baseline throughout the 13.1~day period of the Halo orbits. The less extensive \emph{(u,v)} coverage is fine for photon ring detection but is inferior to that generated by the Earth configuration for a 7~day observational period. Image reconstruction with only this two-element space interferometer would not be possible but the data could still be used to conduct geometric modelling of the source morphology, thus inferring important parameters such as the brightness temperature.

\begin{table}[h!]
    \centering
    \caption{THEZA EML2 orbit configuration for observations at different frequencies.}
    \begin{tabular}{m{3cm} m{3cm} m{3cm} m{3cm}}
        \toprule
        Frequency \newline [GHz] & z-excursion \newline [km] & x-Excursion \newline [km] & Period \newline [days]\\
        \midrule
        590	& 60000  &  25600 & 11.2\\
        690	& 51300  &  22700 & 13.1\\ 
        1200 & 29500 &  16100 & 14.8\\ 
        \bottomrule
    \end{tabular}
    \label{tab:EML2Variation}
\end{table}

\begin{figure}[t]
  \centering
  \includegraphics[width=0.8\columnwidth]{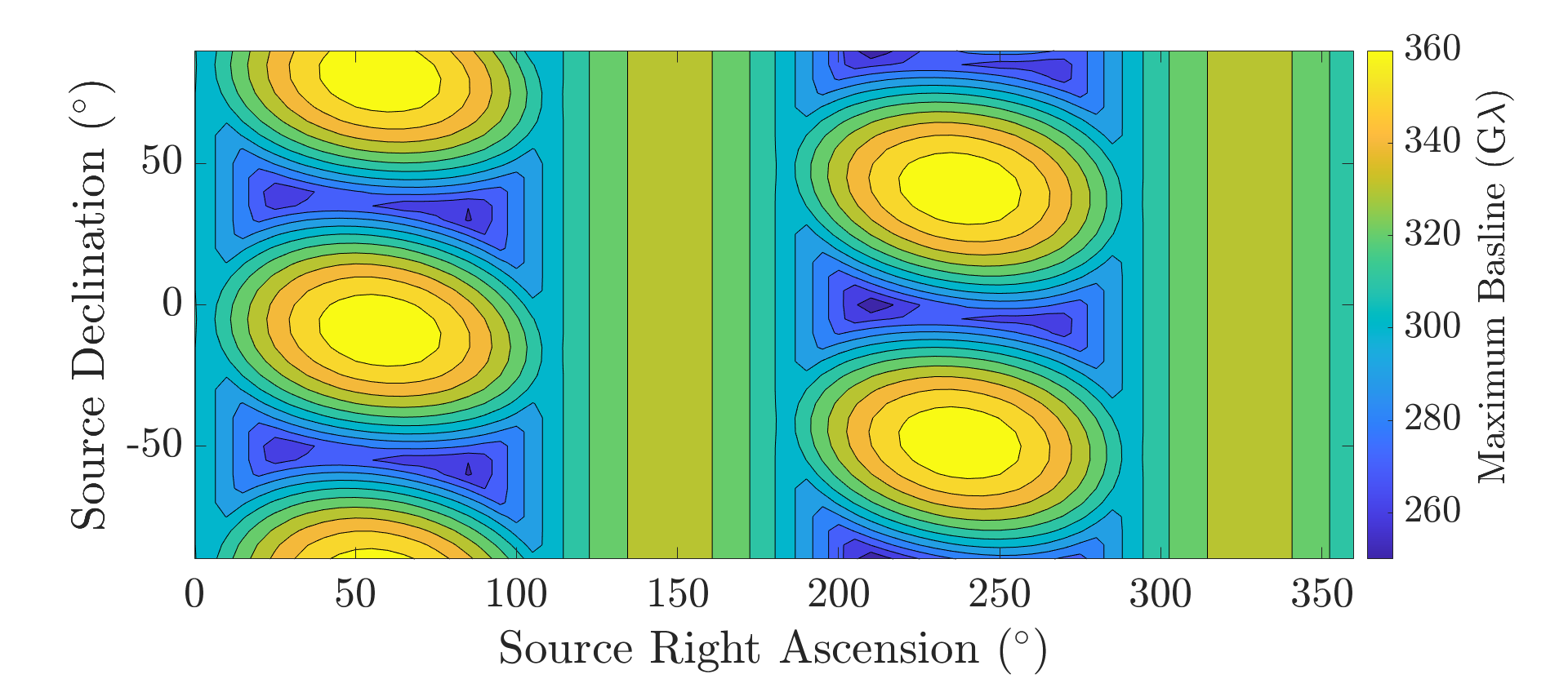}
  \caption{Phase space plot of the maximum baseline length achieved by THEZA in EML2 configuration for sources at all right ascensions and declinations on the celestial sphere.}
  \label{f:EML2phase}   
\end{figure}

The baseline variation required for photon ring detection is achievable for sources at all right ascensions and declinations for observational periods starting at any time during a lunar month. In this Halo configuration, the two elements of the interferometer would pass very close together at the intersection point between the orbits, achieving the shortest baseline. Fig. \ref{f:EML2phase} depicts a phase space plot showing the maximum baseline achieved when observing different sources across a 7~day period. Fig. \ref{f:EML2phase} shows that the baseline length of 200 G\(\lambda\) (i.e. the finest resolution of 1~\(\mu\)as) is achieved for sources at all right ascensions and declinations. In terms of the instrument cadence, with a Nyquist frequency requirement of sampling every 2.5~G\(\lambda\), the interferometer must conduct an observation every 7020~s, for an observational period of 7 days.

Whereas in Earth orbit the planet's thermal emission is not negligible, at the EML2 point the heat flux from combined Earth emission and albedo effect is only 0.13~W/m\(^2\). The Moon albedo effect must instead be considered. At the closest approach the THEZA interferometers in this configuration have a perilune radius of $\sim$18,000~km. This results in a worst case total heat flux incident on the spacecraft of 0.001~W/m\(^2\), assuming an average Moon albedo coefficient of 0.12. The EML2 configuration is therefore favourable when considering the thermal requirements of such a mission.

Conducting a VLBI mission at the EML2 point has a number of additional impacts on the operations. The use of large near-rectilinear Halo orbits (NRHOs) means that eclipses behind the Moon are not experienced at any point in the interferometer's orbits. This enables the possibility of uninterrupted communication with the Earth. As was described in section \ref{ss:earth}, observations by this configuration are also always conducted away from the Sun. Considering this and the lack of eclipses in Halo orbits, the rear of the spacecraft will always be in sunlight providing a constant source of power through the use of large solar panels. This will no doubt be a further enabler of the mission as observation would not need to be halted for periods of time when power generation is too low.

Operation at EML2 does however drastically increase the distance from the Earth. This has implications on the rate of data transmission that is achievable for downlink. This particular challenge is discussed in more detail in section \ref{ss:designVariations}.

\subsection{Sun-Earth L2}
\label{ss:sel2}

\noindent Sun-Earth L2 (SEL2) has long been a favoured position for space missions conducting astronomical observations at a variety of wavelengths. The space telescopes JWST, Planck and Herschel all operate(d) at SEL2 to name a few \cite{rigby_science_2012}. The desirable feature of this orbit is the near-constant Sun position with respect to the periodic orbits, regardless of the time of year. The thermal flux from the Earth is also negligible at this distance. Furthermore, eclipses in orbits around SEL2 are very rare and can be completely avoided when operating in Halo orbits. The L2 point in the Sun-Earth system is located approximately 1.5 million~km from the Earth.

\begin{figure}[t]
\centering
  \includegraphics[width=0.6\columnwidth]{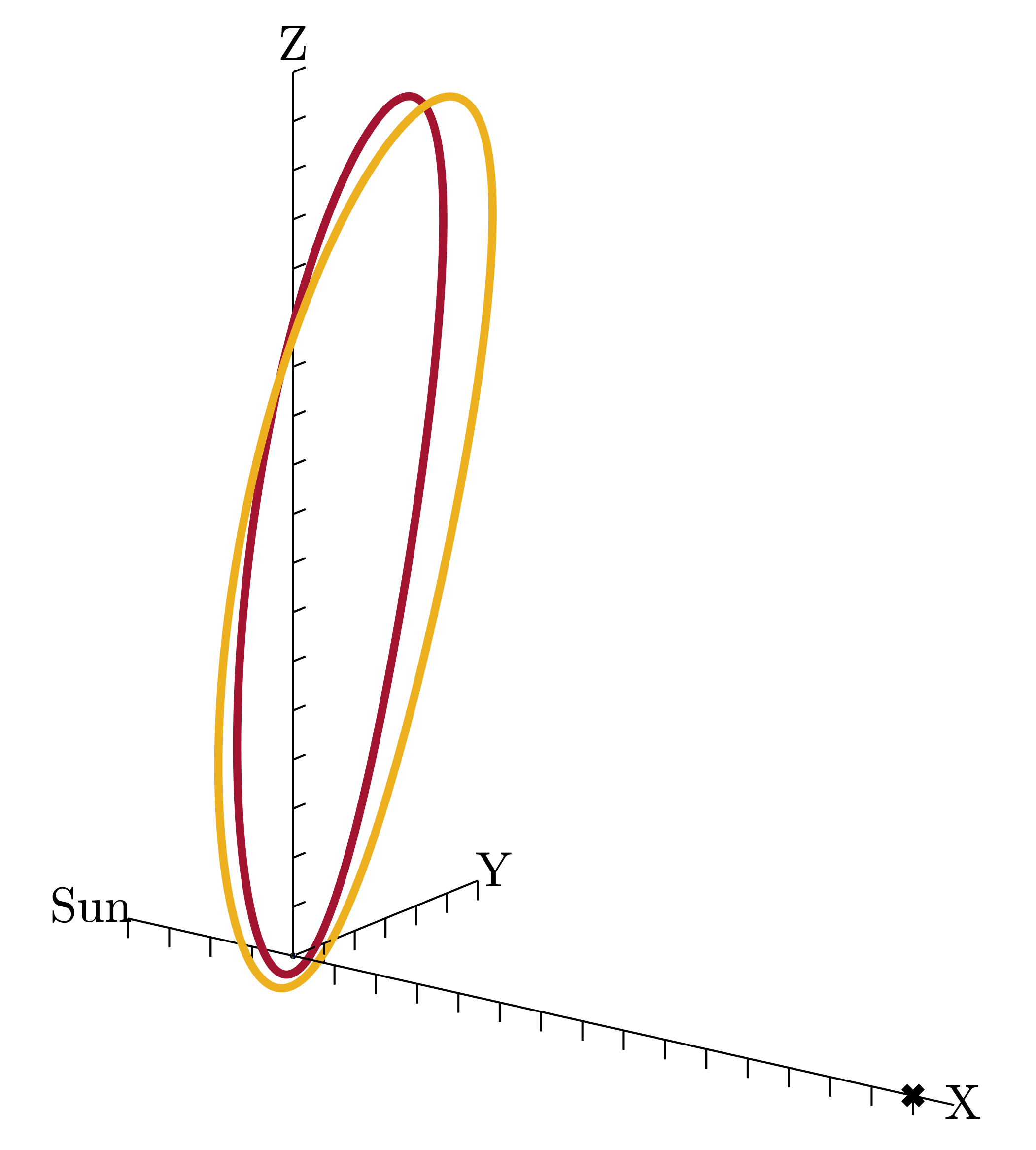}
    \caption{THEZA configuration in SEL2 NRHOs in Earth-centred coordinate system. Observation at 690 GHz. Orbital periods of 90~days (red) and 87~days (yellow). The position of the L2 point is shown by a cross. Tick marks indicate 100,000~km increments.}
  \label{f:SEL2orbit}   
\end{figure}

\begin{figure}[ht]
  \centering
  \includegraphics[width=0.5\columnwidth]{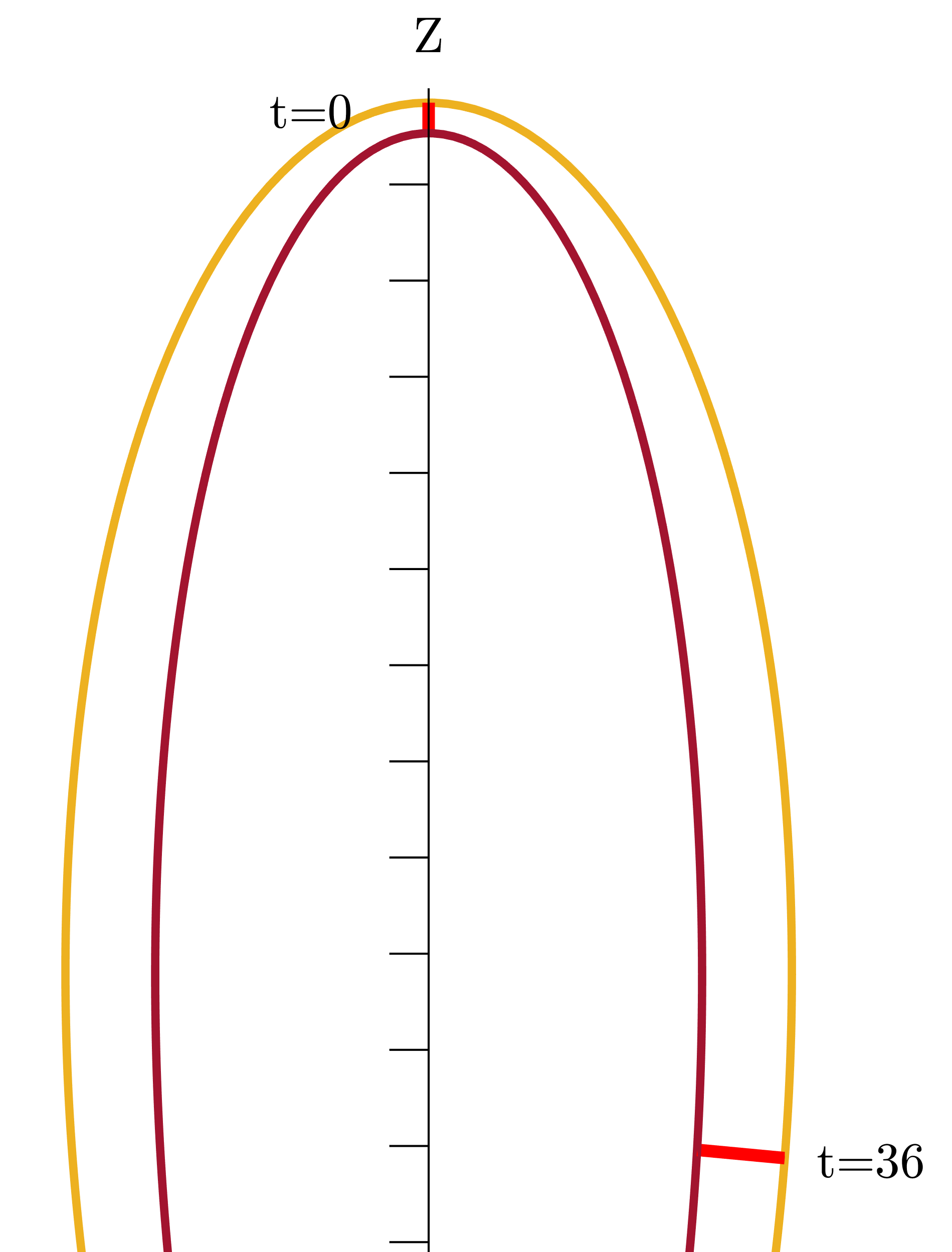}
    \caption{Maximum and minimum length baselines formed by THEZA on SEL2 orbits, projected in the direction of M87*. Tick marks indicate 100,000~km increments.}
  \label{f:SEL2M87View}   
\end{figure}

For conducting VLBI detection of photon rings, the primary disadvantage of operation around SEL2 is the minimum size of the periodic orbits. The JPL Three-Body Periodic Orbit catalogue \footnote{Available at: \url{https://ssd.jpl.nasa.gov/tools/periodic_orbits.html}, accessed on 2023.05.06.} has been used to analyse the families of Halo, Lissajous and vertical Lyapunov orbits that exist. The smallest Halo orbits in the SEL2 system have $z$-excursions of $\sim$400,000~km. As the minimum and maximum baseline length for observation at 690~GHz is 8.7x\(10^3\) and 8.7x\(10^4\)~km respectively, such a large orbit makes use of a Halo geometry like the one presented for the EML2 system impossible.

An alternative geometry utilising two similar Halo orbits could provide the required baseline variation. However, the long periods of SEL2 orbits means that the availability of THEZA to conduct photon ring detection would be severely reduced when compared to the alternative Earth and EML2 configurations. For the system to exhibit a reasonable availability, an NRHO configuration is considered here using Halo orbits with periods of 90 and 87 days. Fig. \ref{f:SEL2orbit} depicts a NRHO configuration in the SEL2 system, for THEZA observations at 690~GHz. The orbits have a close approach to the Earth of 36,000 and 46,000~km, respectively. The spacecraft have an apogee distance of \(\sim\) 1.8~million km. The two orbits have been transformed from an L2-centred frame to an Earth-centred coordinate system for the figure.

Although the period of these orbits is 87 and 90~days respectively, the required baseline variation is achieved in 36~days for M87* and Sgr\,A*. Fig.~\ref{f:SEL2M87View} depicts the maximum and minimum baselines projected in the direction of M87* with the associated time in days at which that baseline is formed.

This analysis is dependent on a specific phasing of the orbits: i.e. both elements of the interferometer cross the Y-Z plane at the same time. As the spacecraft complete multiple revolutions of the orbit, they will slowly drift out of phase increasing the minimum baseline length. Across a period of 365 days, the minimum baseline length would have increased to 70~G\(\lambda\). Although this shortens the baseline range that can be used for photon ring detection, the first order ring is still the dominant contribution to the signal up to \(\sim\)100~G\(\lambda\). Therefore, first order photon ring detection could be performed for the first year of operations. The rest of the mission could then consist of observing on very long baselines for second order ring detection or in collaboration with ground-based arrays to achieve denser \emph{(u,v)} coverage for image reconstruction of the overall source. Although this concept of operations still enables photon ring detection, it is less favourable than the Earth orbit or EML2 configurations which could conduct photon ring detection for the entire mission.

\begin{figure}[t]
  \centering
  \includegraphics[width=0.7\columnwidth]{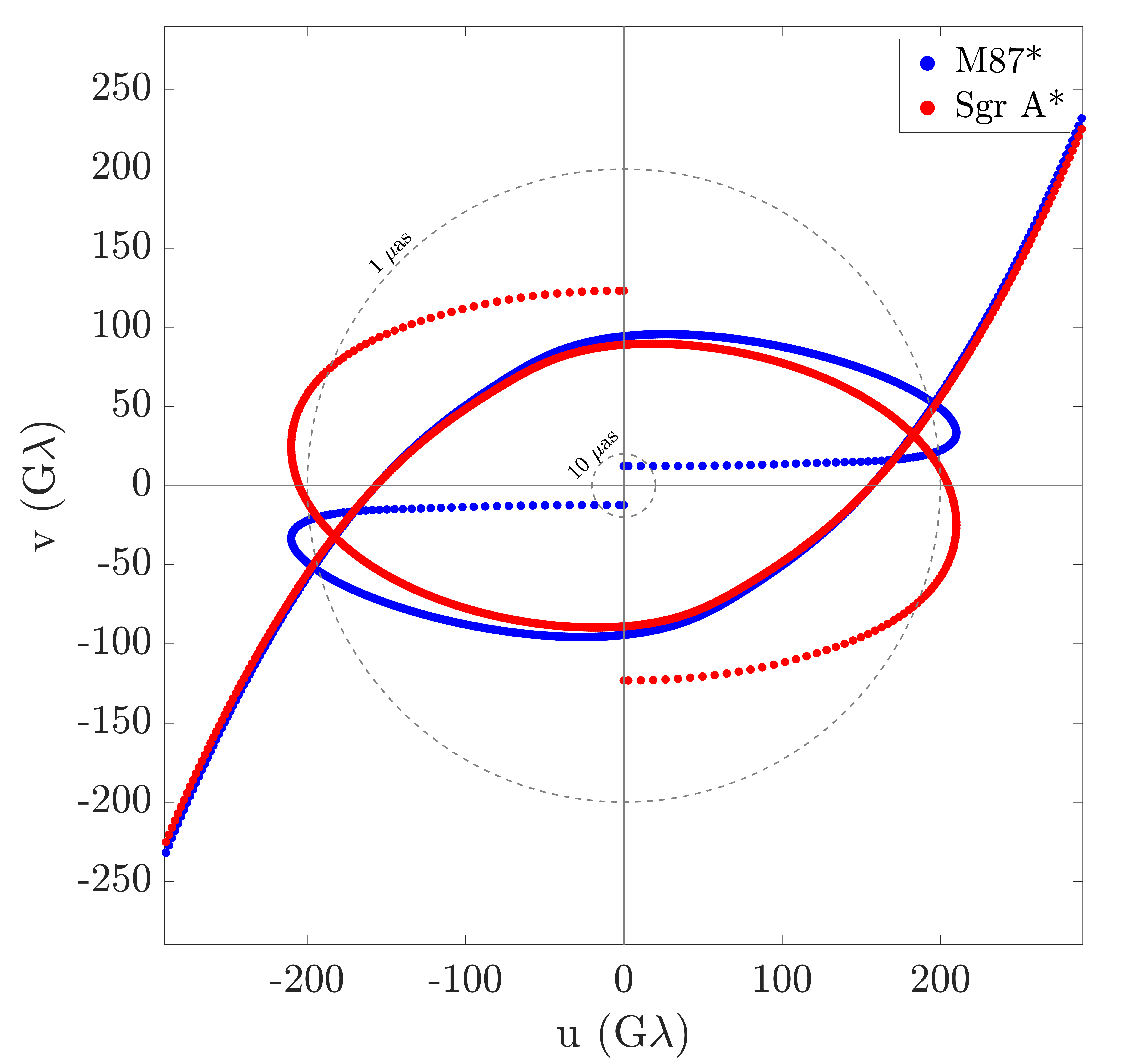}
    \caption{The maximum \emph{(u,v)} coverage of M87* and Sgr\,A* at the time of year when the source is perpendicular to the Y-Z plane. An instrument cadence of 12~hours is used to achieve the required Nyquist sampling of the interferometric signature.}
  \label{f:SEL2uv}   
\end{figure}

As with the Earth-orbiting and EML2 configurations, by always observing away from the Sun individual sources will be available for observations for two, 60~day periods each year. Fig.~\ref{f:SEL2uv} depicts the maximum \emph{(u,v)} coverage of M87* and Sgr\,A* achieved by this THEZA configuration. The longest baseline corresponds to an angular resolution of 0.069~\(\mu\)as (2907~G\(\lambda\)), when observing at 690~GHz. This is a drastic improvement on not only Earth-based instrument resolution but also that which could be offered by the Earth orbit and EML2 configurations of THEZA.

Due to the length of time it takes to achieve the required baseline variation, the instrument cadence of THEZA operating in this configuration is the lowest of the three proposed concepts. An observation would have to be conducted every 12 hours, across the 36 day observational campaign to sample the \emph{(u,v)} plane at the minimum Nyquist frequency. For photon ring detection this length of time does not affect the accuracy of the measured signal. Johnson et al. demonstrate how the visibilities of a time-averaged image are approximately equal to the time-averaged visibilities of a variable image introducing the possibility of utilising long, coherent averaging to conduct photon ring detection \cite{johnson_universal_2020}.

The SEL2 alternative does however have the potential to offer the benefits of the EML2 concept, with regards to thermal considerations, whilst also exhibiting many of the advantages of the Earth-orbiting configuration.

Although at their perigee the spacecraft would pass within $\sim$36,000 and $\sim$46,000~km of the Earth respectively, VLBI observations could be conducted during periods when the interferometer is sufficiently far from the Earth. Considering Fig.~\ref{f:SEL2M87View}, photon ring detection would begin at the $t=0$ position and continue until $t=36$~days. The interferometers would be no closer than 900,000~km from the Earth during imaging, far enough away to make the Earth heat flux negligible. Sufficient time would pass between the perigee position and the start of observations for the telescopes to cool down. 

The close passage to the Earth also offers significant advantages from a communications perspective. A far higher data rate would be achieved during the periods of the interferometer's orbits when the spacecraft are within the geostationary altitude. The use of an optical communication system as described in section \ref{ss:earth} could enable rapid downlink of correlated data. However, this particular configuration would require onboard storage of said data as observation would be conducted 900,000~km from the Earth. This presents its own challenges due to the high level of memory required, and the regular passage through regions of intense radiation.

\section{Discussion}
\label{s:discuss}

\subsection{Photon Ring Detection}
\label{ss:discussPhoton}

\noindent All of the proposed configurations achieve the 20--200~G\(\lambda\) baseline variation required to detect the photon ring interferometric signature. This would enable the primary science goal to be met, resulting in a characterisation of photon rings in M87* and Sgr\,A*, and enabling uniquely robust and accurate tests of strong gravity. Although M87* and Sgr\,A* are considered as the primary targets of THEZA, each of the proposed configurations can achieve the baseline variation (and thus, angular resolutions of between 10~\(\mu\)as and 1~\(\mu\)as) for radio sources at all positions on the celestial sphere. Such a system could therefore conduct a comprehensive survey of supermassive black holes in active galactic nuclei. The combination of extreme angular resolution and high observing frequency would enable the optically thin cores in a number of objects to be resolved, providing new mass estimates of their central black holes (e.g., \cite{johnson_key_2023}). 

Operating in any of the three proposed configurations, with an integration time of 1000~s and 32~GHz bandwidth, THEZA would achieve sufficient sensitivity to enable characterisation of the first order photon ring. In order to accurately estimate the diameter of the \(n=2\) ring at \(\sim\)200~G\(\lambda\), a sensitivity of \(\sim\)0.2~mJy would likely be required, a factor of 2 lower than achievable with this THEZA configuration, operating at 1000~s and with a 64~GHz bandwidth. However, conducting multiple observations at the same \emph{(u,v)} points could enable the use of incoherent stacking of the measurements to effectively reduce the error in the ring diameter estimation. This method would theoretically enable detection of the second order ring.

\subsection{VLBI Imaging}
\label{ss:discussImaging}

\begin{figure}[t!]
  \centering
  \includegraphics[width=0.7\columnwidth]{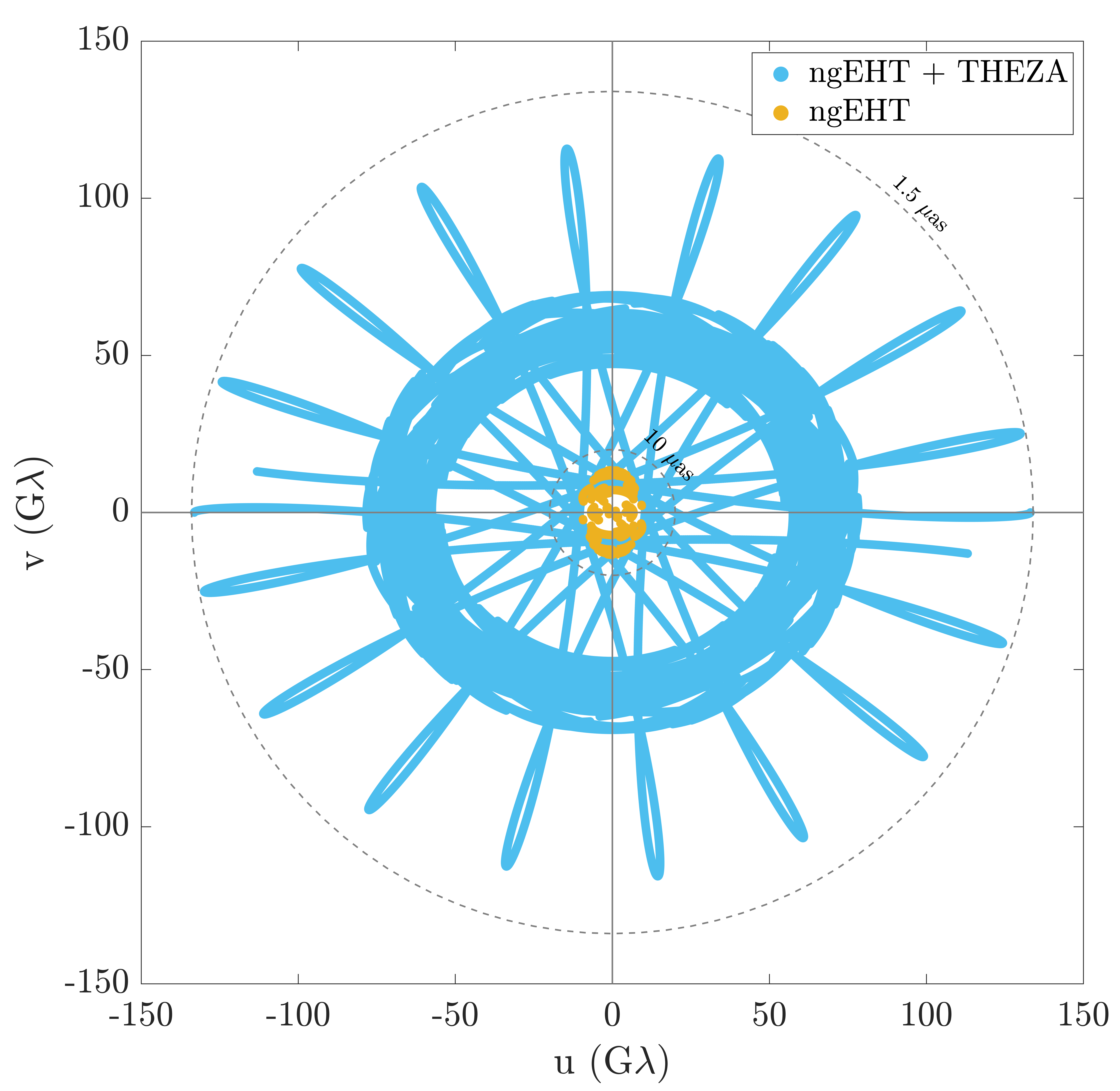}
    \caption{\emph{(u,v)} coverage of M87* by THEZA and the current configuration of the EHT observing at 345~GHz, the target frequency of the ngEHT \cite{blackburn_studying_2019}.}
  \label{f:ngEHTuv}   
\end{figure}

\noindent For a mission of the magnitude discussed here, the system would need to be highly versatile in order to justify the investment required. Conducting VLBI observations for general image reconstruction of the overall source would also be necessary. As described in section \ref{ss:VLBI}, optimising \emph{(u,v)} coverage is essential for high-fidelity image reconstruction. The radio emitting environment of black holes is also highly variable over short timescales, comparable to the light cross-section time of Schwarzschild radii. In order to capture the dynamic behaviour of the SMBHs, the required \emph{(u,v)} coverage should be achieved in as short a time as possible (see \cite{roelofs_ngeht_2023} for relevant discussion). Considering the various configurations proposed here, for VLBI observations on space-space baselines the Earth orbiting proposal provides the most rapid and dense \emph{(u,v)} coverage. The \emph{(u,v)} coverage achieved by the EML2 and SEL2 are far more sparse and would likely be insufficient for image reconstruction when operating as only a two-element, space-based array.

However, implementing multiple receivers onboard the spacecraft for observations at different frequencies would enable THEZA to operate in collaboration with ground-based arrays. The multi-frequency capability would also enable very valuable phase calibration by means of frequency phase transfer methods \cite{rioja_transformational_2023}. The required geometry for photon ring detection could be set based on the higher frequency utilised and a lower frequency receiver would be used to conduct ground-space VLBI. Fig. \ref{f:ngEHTuv} illustrates the \emph{(u,v)} coverage achievable across a 7-day observation period with the EHT and THEZA in the Earth orbit configuration, at 345\,GHz the planned frequency of the ngEHT \cite{blackburn_studying_2019}. This figure clearly depicts the improvement in \emph{(u,v)} coverage and angular resolution provided by a two-element, spaceborne VLBI configuration.

The primary metric of the performance of the photon ring detection application means that VLBI measurements can be conducted across a large range of spatial scales. Ranging from \(\sim\)20 to 200~G\(\lambda\), the proposed configurations would make THEZA sensitive to structures in the source of a variety of sizes. THEZA would be capable of detecting the smallest structures in sources with the maximum baseline length of 200~G\(\lambda\) corresponding to an angular resolution of 1~\(\mu\)as.

\subsection{Future Technologies}
\label{ss:designVariations}

\noindent The system architecture and high-level spacecraft design considered in this concept is described in section \ref{s:system}. Some elements of this concept have been derived from higher requirements (such as the primary frequency selection) whereas others have been postulated in order to provide a baseline which could be used for preliminary analysis. A number of intriguing variations in the proposed system design and operation are summarised in this section. With sufficient development in key technologies, these design features have the potential to dramatically impact the capabilities of a space-based VLBI concept.

Gurvits et al. \cite{gurvits_science_2022} note that in-orbit, real-time correlation of data between the two interferometer elements using an Inter-Satellite Link (ISL) can drastically reduce the required downlink data rate. For an interferometer operating at 64~Gbps, a correlator producing 8-byte complex numbers could reduce the total data rate to 262~Mbps. Reduction of the data rate to this level would not only simplify data handling and communications for Earth orbit configurations but would be achievable at Lunar distances using a traditional, high frequency radio communication. RadioAstron downlinked data at 144~Mbps in the Ku-Band at Earth-Moon distances \cite{kardashev_radioastron-telescope_2013}. With an increased antenna diameter and/or transmit power, 262~Mbps would be feasible, overcoming a key challenge of the EML2 configuration. Performing in-orbit correlation would however require sufficient onboard computing power and a very high data rate, optical ISL. A trade-off between performing real-time correlation and post-observation processing would also have to be resolved. The former would require maintenance of the ISL link throughout observations, most likely limiting the times at which observation of certain sources could be performed. The latter would require high volume data storage systems and also impact the operations of the mission as time would need to allocated to perform the correlation. Despite the difficulties, in-orbit correlation could provide an interesting means of overcoming the communications and data handling challenges that space-based VLBI faces.

\begin{figure}[t!]
  \centering
  \includegraphics[width=0.8\columnwidth]{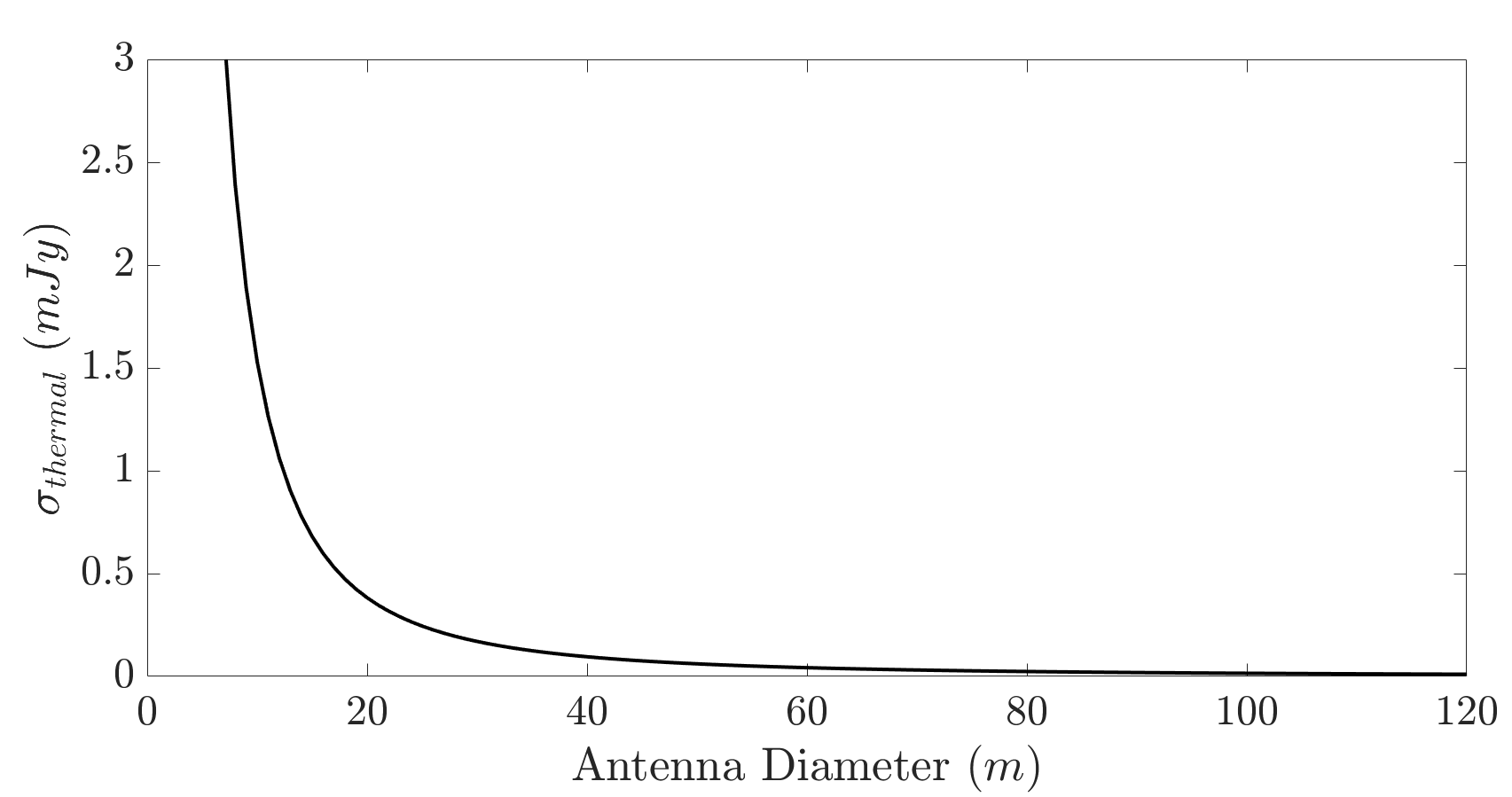}
    \caption{Thermal noise variation with antenna diameter for a bandwidth of 32\,GHz and \(t_{int}=1000s\).}
  \label{f:antennaThermal}   
\end{figure}

As was discussed in section \ref{s:photonRingDetection}, sensitivity of the interferometer directly impacts the ability to conduct photon ring detection. As well as increasing integration time and bandwidth, sensitivity can also be improved by utilising an antenna with a larger collecting area as depicted in Fig.~\ref{f:antennaThermal}. Moving beyond the 15~m diameter deployable antenna proposed here, the possibility of future in-orbit assembly of phased array systems and launch vehicle developments could drastically improve the accuracy of measurements. An antenna with a diameter of 40~m would improve the baseline sensitivity of the system to 0.1~mJy, sufficient to characterise the \(n=2\) photon ring without incoherent averaging.

\section{Conclusions}
\label{s:concl}

\noindent Three intriguing  configurations have been proposed for a THEZA mission conducting space-based VLBI observations to study the photon ring phenomenon. Each configuration has undergone a preliminary evaluation to assess its ability to successfully service this application.

As well as performing a primary mission of photon ring detection, the THEZA configurations proposed could be used for achieving other science goals. VLBI imaging of SMBH targets could be conducted in the same way as the EHT by implementing multiple receivers onboard, operating at different frequencies. On purely space-space baselines, THEZA would achieve an angular resolution of up to 1~\(\mu\)as, meeting the original aim of providing an order of magnitude improvement over current EHT and ngEHT capabilities. In particular, the Earth-orbiting configuration can achieve far greater \emph{(u,v)} coverage than the EHT across the same, 7-day observational time period. Performing observations in conjunction with ground-based antennas, the generated images would have higher fidelity and include detail of far smaller structures than the EHT, or the proposed ngEHT, due to the improvement in angular resolution.

Key technical challenges related to the implementation of a mm-wavelength, space-based VLBI system still exist: baseline determination, data handling and thermal control to name a few. Although high-level solutions to some of the problems have been discussed here and in other studies (see \cite{gurvits_science_2022, roelofs_simulations_2019, kurczynski_event_2022}), significant research and technological development is required to overcome these difficulties. Despite the challenges, the scientific outputs of such a mission would be highly valuable to the community.

Due to the inherent limitations of Earth-based radio astronomy, in order to achieve the objectives of THEZA a space-based system is unavoidable. It is hoped that this work can serve as a starting point for future, more detailed mission concept design and analysis for a THEZA system studying photon rings around SMBHs and conducting general VLBI imaging of astronomical radio sources.

\section*{Acknowledgements}
\noindent The authors appreciate very valuable comments and suggestions made by five anonymous referees. The authors are grateful to Michael Janssen for his comments on sections \ref{ss:VLBI}, \ref{ss:science} and \ref{s:photonRingDetection}. WZ and LL acknowledge support of the National Natural Science Foundation of China (Grant No. 11973011) and the Key Incubation Project of Shanghai Astronomical Observatory.

\bibliographystyle{elsarticle-num} 
\bibliography{bibliography}

\end{document}